\documentclass{aa}  
\usepackage{txfonts}
\usepackage{graphicx}
\usepackage{mathrsfs}
\usepackage{siunitx}
\usepackage{hyperref}
\usepackage{soul}
\usepackage[dvipsnames]{xcolor}
\usepackage{comment}
\usepackage{siunitx}
\usepackage{float}
\usepackage{pifont}

\newcommand{\labelization}[1]{{\fontfamily{cmtt}\selectfont #1}}

\begin{document}

\defcitealias{Wafflard-Fernandez&Lesur2023}{WFL23}

\title{Gone with the wind: the outward migration of eccentric giant planets in windy disks}
\titlerunning{Outward migration of eccentric giant planets in wind-emitting disks}
\author{
    Gaylor Wafflard-Fernandez\inst{\ref{i:ipag}}%
    \thanks{\email{gaylor.wafflard@univ-grenoble-alpes@fr}}
\&
    Geoffroy Lesur\inst{\ref{i:ipag}}
}
\institute{
    Univ. Grenoble Alpes, CNRS, IPAG, 38000 Grenoble, France\label{i:ipag}
}
\authorrunning{GWF \& GL}

\date{Received xxx; accepted xxx}

\abstract
{Recent studies indicate that circumstellar disks exhibit weak turbulence, with their dynamics and evolution being primarily influenced by magnetic winds. However, most numerical studies have focused on planet-disk interactions in turbulent disk models.}
{We aim to explore how wind-driven accretion affects the orbital and eccentricity evolution of a Jovian planet within a magnetized disk. Conversely, we seek to determine in what extent such a planet can modify the accretion behavior and the wind dynamics.}
{We perform high-resolution 3D global non-ideal magneto-hydrodynamic (MHD) simulations of a massive gap-carving planet interacting with a wind-launching disk, using the accelerated code \labelization{IDEFIX}. We consider the influence of the gap shape on planet migration by restarting a "fixed-planet" simulation at three different times, from which the planet evolves freely in the disk.}
{For a strong initial magnetization and a sufficiently deep planet gap, we find that the planet becomes moderately eccentric, and its migration is slow, unsteady and mostly outward. This migration pattern is due to the gap's radial asymmetry which enhances the inner Lindblad torque while reducing the outer Lindblad torque. We show that eccentricity can grow up to $6-8\%$ and is likely driven by a finite-amplitude instability triggered by first-order external Lindblad resonances. These moderate eccentricity values periodically modulate the gap accretion rate and wind mass loss rate, possibly leading to the formation of discrete structures in CO outflows.}
{Slow outward migration and eccentricity growth appear to be common outcomes of planet-disk-wind interactions, which may contribute significantly to both the long orbital periods and the moderate eccentricities of warm jupiters. Additionally, eccentric massive protoplanets embedded in circumstellar disks could play a role in generating structured outflows.}


\maketitle 

\section{Introduction}

About one-third of giant planets detected around evolved stars, with masses between that of Saturn and 5 times that of Jupiter, lie on relatively large orbital radii beyond $1$ au. Among them, nearly two-thirds are moderately eccentric in the range $[5\%-40\%]$ (data taken from \href{https://exoplanet.eu/catalog/}{exoplanet.eu}). The orbital properties of these numerous so-called warm jupiters remain elusive, and raise the question of their origin. Because these planets must have formed in a gas-rich environment \citep[see, e.g.,][]{Drazkowska2023}, they are likely to have experienced gravitational interaction with a massive protoplanetary disk, altering their orbital distance via planetary migration \citep[see, e.g.,][]{Paardekooper2023}.

The migration of giant planet has been extensively studied since the trailblazing work by \cite{Lin&Papaloizou1986c}. In a nutshell, a giant planet that is massive enough to perturb the gas density profile depletes its co-orbital region \citep{Lin&Papaloizou1986b}. This non-linear mechanism separates the disk in an inner and an outer region. In a viscous disk model, the disk has a finite lifetime and is therefore expected to evolve and lose its material on a viscous timescale. This accretion process refills partially the gap as matter is accreted inwards, carrying the gap-carving planet with it. This is the classical picture of type-II migration in viscous disks, where massive planets are locked within their gaps. Later works have shown that the gap does not completely isolate the outer disk from the inner disk and gas may actually cross the gap region \citep{Duffell2014}, leading to a migration rate different from the viscous accretion rate, but still proportional to the viscosity \citep{Robert2018}. For a simple gas surface density power-law profile $\Sigma=89~{\rm g.cm}^{-1}\left(r/10~{\rm au}\right)^{-0.5}$ and a moderate accretion rate $\dot{M}=10^{-8}M_\odot$.yr$^{-1}$, a planet located at $10~\rm au$ will migrate inwards, with a radial speed of $\sim16$ au.Myr$^{-1}$. Without gravitational scattering events by one or several companions \citep{Chatterjee2008}, the eccentricity of such single planet is classically expected to be damped by planet-disk interaction processes \citep{Bitsch2013}. Within the viscous accretion framework and for a realistic stellar accretion rate \citep{Venuti2014}, massive planets should therefore migrate inwards on a timescale much shorter than the disk lifetime and eventually pile up on very short-period circular orbits \citep[see, e.g., the review on hot jupiters by][]{Dawson&Johnson2018}, which makes the existence of eccentric warm jupiters puzzling.

One way of alleviating this tension is to question the physical processes that drive accretion. In the viscous scenario, accretion results from a radial transport of angular momentum by the viscous radial stress. This viscosity stands as an effective model for the small scale (i.e., with eddies smaller than the disk scale height) turbulence. The best candidate to explain this turbulence is the magneto-rotational instability \citep[MRI,][]{Balbus&Hawley1991,Lesur2023a} However, theoretical analysis and simulations show that the MRI -- and thus the resulting turbulence -- should be severely quenched by non-ideal MHD effects in the outer regions of the disks \citep[see, e.g.,][]{Perez-Becker&Chiang2011a,Perez-Becker&Chiang2011b,Bai&Stone2013,Lesur2014}, where warm jupiters are expected to evolve. In addition, the radio-interferometer ALMA (Atacama Large Millimeter Array) reveals a collection of thin dust substructures in the radio continuum emission \citep[see, e.g.,][]{Andrews2018}, which also suggests a very low level of turbulence \citep{Pinte2016,Villenave2020}. These theoretical and observational constraints, together with the increasing number of detections of CO outflows (see, e.g., \citealt{Louvet2018,deValon2020,deValon2022,Pascucci2023}), have revived an alternative scenario for accretion: a magnetic wind extracts the disk's angular momentum \citep{Blandford&Payne1982,Bai&Stone2013}, and matter accretion in the radial direction is essentially a laminar process \citep{Wardle1993}. 

An important outcome of this magnetized wind scenario is that high accretion can be sustained with a low level of turbulence, which could significantly reduce inward migration of massive planets and therefore help the survival of long-period warm jupiters. Recently, various numerical studies have focused on assessing the impact of these magnetized winds on planet-disk interactions: (i) In low-viscosity disks with prescribed angular momentum removal, \cite{Kimmig2020} and \cite{WuChen2025} have found with 2D hydrodynamic simulations that giant planets can undergo episodes of runaway type-III-like outward migration \citep[see, e.g.,][]{Masset2003,Peplinski2008c} if the wind mass loss rate is high enough. \cite{Lega2021,Lega2022} have conducted purely hydrodynamic 3D simulations and have obtained an inward vortex-driven migration that eventually slows down, or is sometimes reversed and stops as long as the accretion flow is not blocked by the planet gap. Such planet remains on a quasi-circular orbit. 
(ii) 3D non-ideal MHD simulations have treated magnetized winds self-consistently by threading the disk in a large-scale weak vertical magnetic field, focusing on the impact of a giant planet on the structures of the gaseous disk \citep[][hereafter \citetalias{Wafflard-Fernandez&Lesur2023}]{Aoyama2023,Hu2025,Wafflard-Fernandez&Lesur2023}. In particular, \citetalias{Wafflard-Fernandez&Lesur2023} have shown that a massive planet in a highly magnetized disk carves a radially asymmetric gap that eventually erodes the outer disk. This mechanism, which also occurs in magnetized inner cavities  \citep{Martel&Lesur2022}, results from a slightly higher accretion rate in the low-density region. Such asymmetry produces a planet gap that is wider than expected, and could be responsible for an outward migration of giant planets \citepalias{Wafflard-Fernandez&Lesur2023}. 
However, all these MHD studies have been performed under the assumption of planets in fixed circular orbits. 

This is precisely the hypothesis that we relax in the present paper. We carry out 3D global non-ideal MHD simulations, and consider in a self-consistent way the impact of magnetized winds on the evolution of a giant planet's orbital properties. In particular, we let a Jupiter-mass planet evolve freely in a circumstellar disk to study its orbital migration and eccentricity evolution, focusing on the impact of the planet gap shape. The paper is organized as follows. We first describe in Section~\ref{sec:numeric} the numerical setup of the MHD simulations. The results are then presented in Section~\ref{sec:3d_results}, with a focus on disk structures in Sections~\ref{sec:disk_structures_variability}, planet's orbital properties in Sections~\ref{sec:planet_migration_eccentricity},~\ref{sec:gravitational_torques},~\ref{sec:main_eccentricity} and~\ref{sec:magnetization}, as well as accretion and wind behavior in Section~\ref{sec:main_modulation}. Finally, we summarize and discuss the implications of our results in Section~\ref{sec:conclusion}.

\section{Numerical methods and setups}
\label{sec:numeric}
\subsection{Code and grid}
We carried out 3D non-ideal MHD simulations using the code \href{https://github.com/idefix-code/idefix}{\labelization{IDEFIX}} \citep{Lesur2023}, which is accelerated by graphics processing units (GPU) and has already been used for various astrophysical problems \citep{VandenBossche2023,Mauxion2024}. For this work, \labelization{IDEFIX} was run both on the Jean Zay cluster (IDRIS) on Nvidia V100 GPUs and on the Ad Astra cluster (CINES) on AMD Mi250X GPUs, using a second order Runge–Kutta scheme, second order spatial reconstruction and the Harten-Lax-van Leer discontinuities (HLLD) Riemann solver. The timestep being limited by the Alfvén velocity, we do not make use of the FARGO orbital advection scheme \citep{Masset2000}. The setup as well as the numerical method that we used are largely described in Section~2 of \citetalias{Wafflard-Fernandez&Lesur2023}. We summarize here briefly some of the main features and how the numerical protocol used in this present study differs from it. 

First, we adopt a spherical coordinate system ($r$, $\theta$, $\phi$), with $r$ the radial spherical coordinate, $\phi$ the azimuthal angle, and $\theta$ the colatitude. In the azimuthal direction, the grid is uniform and extends from $0$ to $2\pi$ with $1024$ cells. In the radial direction, the grid is logarithmic between $r_{\rm{min}}=0.08r_{\rm{p,0}}$ and $r_{\rm{max}}=5r_{\rm{p,0}}$ with $512$ cells, where $r_{\rm{p,0}}=1$ is also the initial location of the planet. In the latitudinal direction, we use a uniform grid around the midplane with $148$ cells between $\pi/2 \pm \arctan{6h}$ and two stretched grids symmetric around that refined region with $54$ cells on both sides and extending up to the axes. The aspect ratio $h_0=c_{s,0}/v_K=0.05$ is constant, $c_{s,0}$ being the locally isothermal sound speed, $v_K=r\Omega_K$ the local Keplerian velocity, and $\Omega_K$ the local Keplerian orbital frequency. The total resolution is $N_r \times N_\theta \times N_\phi = 512 \times 256 \times 1024$, and corresponds respectively for the radial, latitudinal and azimuthal directions to $\simeq14$, $13$ and $8$ points per disk pressure scale height $H(r) = h_0r$ around the planet location. To avoid too small cells around the polar axis, we use a ``grid coarsening`` procedure (also known as ring average)  that effectively increases the azimuthal size of cells close to the axis, following the procedure of \cite{Zhang2019}.

\subsection{Equations of motion, non-ideal effects}
\labelization{IDEFIX} solves the non-ideal MHD equations for the density $\rho$, the velocity field $\vec{v}$ and the magnetic field $\vec{B}$. We include ambipolar diffusion in the disk bulk (see below) but neglect the Hall effect. Ohmic resistivity is enabled only close to the radial boundaries to create damping buffer.  As in \cite{Lesur2021}, we prescribe the ambipolar diffusion via the dimensionless ambipolar Elsässer number
\begin{equation}
    \displaystyle A_{\rm m} = \max\left\{\tilde{A}_{\rm m} \exp{\left[\left(\frac{z}{\epsilon_{\rm id} H(R)}\right)^4\right]},\left(\frac{v_A}{c_{s,0}}\right)^2\right\},
    \label{eq:am}
\end{equation}
with $\tilde{A}_{\rm m}=1$ the midplane Elsässer number, $v_A = B/\sqrt\rho$ the Alfvén speed and $\epsilon_{\rm id}=6.0$. This prescription in Eq.~\ref{eq:am} helps mimic the ionization by X-rays and far UVs at the disk surfaces expected from complex thermo-chemical models \citep{Thi2019} while saving computational time. More elaborate models with self-consistent thermochemistry of a wind-launching non-ideal MHD disk with gaps show similar results \citep{Hu2023}

\subsection{Initial conditions and cooling}

The initial condition is a disk in hydrostatic equilibrium threaded by a large-scale vertical magnetic field set initially to have the plasma parameters $\beta_0=10^3$ or $\beta_0=10^5$ in the disk midplane, following a vector potential approach as in \cite{Zhu&Stone2018}. For the gas surface density at $r_{\rm{p,0}}=1$, we choose $\Sigma_0=10^{-3}$ in code units, which corresponds for a solar-mass star to $88$ g.cm$^{-2}$ at $10$ au, and a radial profile following $\Sigma_{\rm{ini}}=\Sigma_0(r/r_{\rm{p,0}})^{-0.5}$. Regarding the temperature stratification, the simulation domain is divided into a dense cold disk and a low-density hot corona, with a smooth transition between these two regions as in \citetalias{Wafflard-Fernandez&Lesur2023} (see their Eq.~19). We choose a fast $\mathcal{B}$-cooling prescription in order for the temperature to quickly relax towards the prescribed profile, using
\begin{equation}
    \displaystyle \mathcal{B}=\left[\tau_{\rm{grain}}+(\tau-\tau_{\rm{grain}})\exp{\left(4\frac{r-r^{\rm{WKZ}}_{\rm{min}}}{r-r_{\rm{min}}}\right)}\right]\Omega_K^{-1},
    \label{eq:bcool}
\end{equation}

\noindent with $\tau_{\rm{grain}}=0.01$ close to the grid's inner edge and $\tau=0.1$ beyond the radius $r^{\rm{WKZ}}_{\rm{min}}=0.15r_{\rm{p,0}}$. This complex $\mathcal{B}$ is a good compromise to quench the vertical shear instability \citep[VSI,][]{Nelson2013, Stoll&Kley2014,Manger2021} and avoid a strong decrease of the timestep due to fast heating of gas material close to the inner boundary. The two regions $r\in[r_{\rm{min}}-r^{\rm{WKZ}}_{\rm{min}}]$ and $r\in[r^{\rm{WKZ}}_{\rm{max}}-r_{\rm{max}}]$, with $r^{\rm{WKZ}}_{\rm{max}}=4.7r_{\rm{p,0}}$, define the wave-killing zones we adopt to prevent reflection of waves near the boundaries. At the inner (outer) radius, we copy most of the fields -- the density $\rho$, the pressure $P$, the velocity components ($v_r$, $v_\theta$, $v_\phi$) and $B_\theta$ -- from the innermost (outermost) active zones to the ghost zones. We make sure that material does not come out from the ghost zones to the active grid by applying an additional condition on $v_r$. For the other fields, we impose $B_\phi=0$ and $B_r$ is reconstructed from the divergence-free condition on $B$. The boundary conditions are periodic in the azimuthal direction, and we apply a regularization procedure around the axis in the latitudinal direction $\theta$ so as to make it transparent to the flow, following the procedure in the appendix of \cite{Zhu&Stone2018}. To avoid too small timesteps, we use the thresholds $v_A<v_{A,\rm{max}}=v_K(r_{\rm{min}})$ and $\rho<\rho_{\rm min}=10^{-12}$, as in \citetalias{Wafflard-Fernandez&Lesur2023}.

\subsection{Planet properties}

For the planet, we adopt a planet-to-primary mass ratio of $q_p=10^{-3}$, which represents a Jupiter-mass planet orbiting a solar-mass star. Accretion on the planet is neglected. The total gravitational potential in the simulations takes into account the potential of the star, the Plummer potential of the planet, and the indirect term arising from the acceleration of the star by the planet. To compute the planet migration and because gas self-gravity is neglected, we remove the material inside $50\%$ of the Hill sphere when computing the gravitation torque exerted by the gas onto the planet, as suggested in \cite{Crida2009}.

\subsection{Average conventions}

For a given quantity $Q$, the azimuthal and latitudinal averages are defined in Eq.~\ref{eq:average_phi} and \ref{eq:average_theta} respectively: 
\begin{equation}
    \langle Q \rangle_\phi = \displaystyle \frac{1}{2\pi} \int_{0}^{2\pi}Q d\phi,
    \label{eq:average_phi}
\end{equation}
\vspace{-0.5cm}
\begin{equation}
    \overline{Q} = \displaystyle \int_{\theta_-}^{\theta_+} Q r\sin{\theta}d\theta,
    \label{eq:average_theta}
\end{equation}

\noindent with Eq.~\ref{eq:average_theta} integrated between angles $\theta_-$ and $\theta_+$ such that
\begin{equation}
    \displaystyle \frac{\theta_+ - \theta_-}{2} = \arctan(3h).
    \label{eq:surface}
\end{equation}

\subsection{Numerical protocol}

Before introducing the planet, we run an axisymmetric simulation with $\beta_0=10^3$ for $\simeq200$ orbits at $r=1$. We also run an axisymmetric simulation with $\beta_0=10^5$ at lower resolution. Once a magnetic wind has been launched and has reached a quasi-steady state, we continue the simulations in full 3D and introduce the planet, positioned at $r_{\rm{p,0}}=1$. This instant marks the $t=0$ of the $3$ main simulations presented in this work. The planet mass is then gradually increased to reach its final mass after $10$ orbits. We keep the planet on a fixed circular orbit up to $t=T_\mathrm{release}$ when the planet is eventually allowed to migrate freely, dynamically responding to the torques exerted by the surrounding gas. Hence, each simulation we present is divided into 3 stages:
\begin{enumerate}
\item Planet introduction for $0<t<10$
\item Planet on a fixed circular orbit for $10<t<T_\mathrm{release}$
\item Planet migrating for $t>T_\mathrm{release}$
\end{enumerate}

\noindent Both the strong accretion variability and the secular evolution of the outer gap make it difficult to determine $T_\mathrm{release}$. In order to tackle this issue, we identify each simulation $\mathcal{M}$ by the instant at which the planet is released: $\mathcal{M}_{10}$ has $T_\mathrm{release}=10$, $\mathcal{M}_{200}$ has $T_\mathrm{release}=200$, etc. Our aim is therefore to determine how the planet-induced gap shape impacts the migration of a Jupiter-mass planet in a strongly magnetized wind-launching disk. Approximately $400\,000$ GPU hours have been used for the main simulations presented here, which corresponds to an estimated carbon footprint of the order of $15$ tons of CO$_2$.

\section{Results}
\label{sec:3d_results}

\begin{figure*}
    \centering
    \includegraphics[width=0.955\hsize]{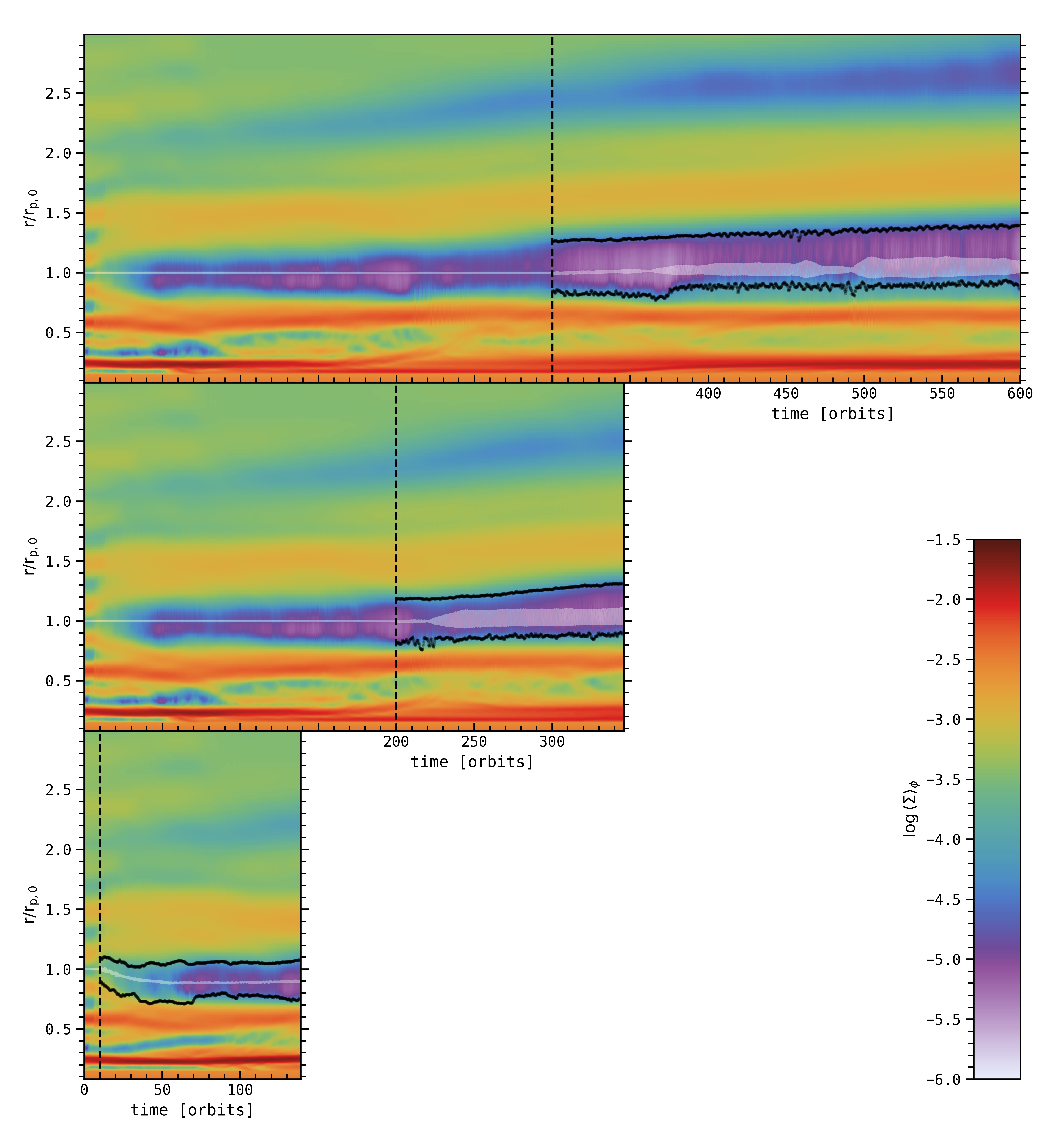}%
    \caption{Space-time diagram showing the evolution of the azimuthally-averaged gas surface density $\Sigma$, for the three runs $\mathcal{M}_{300}$ (top panel), $\mathcal{M}_{200}$ (middle panel) and $\mathcal{M}_{10}$ (bottom panel). The radial extent of all the diagrams range from $0.08r_{\rm{p,0}}$ to $3r_{\rm{p,0}}$, with the planet initially at $r_{\rm{p,0}}=1$. The fixed regime is separated from the migrating regime by the vertical black dashed line, and the instantaneous position of the planet is represented by the semi-transparent white line. The estimated inner and outer gap edges in the migrating regime are indicated by the black dotted lines.}
    \label{fig:main_spacetime}
\end{figure*}

\begin{figure*}
    \centering
    \includegraphics[width=0.955\hsize]{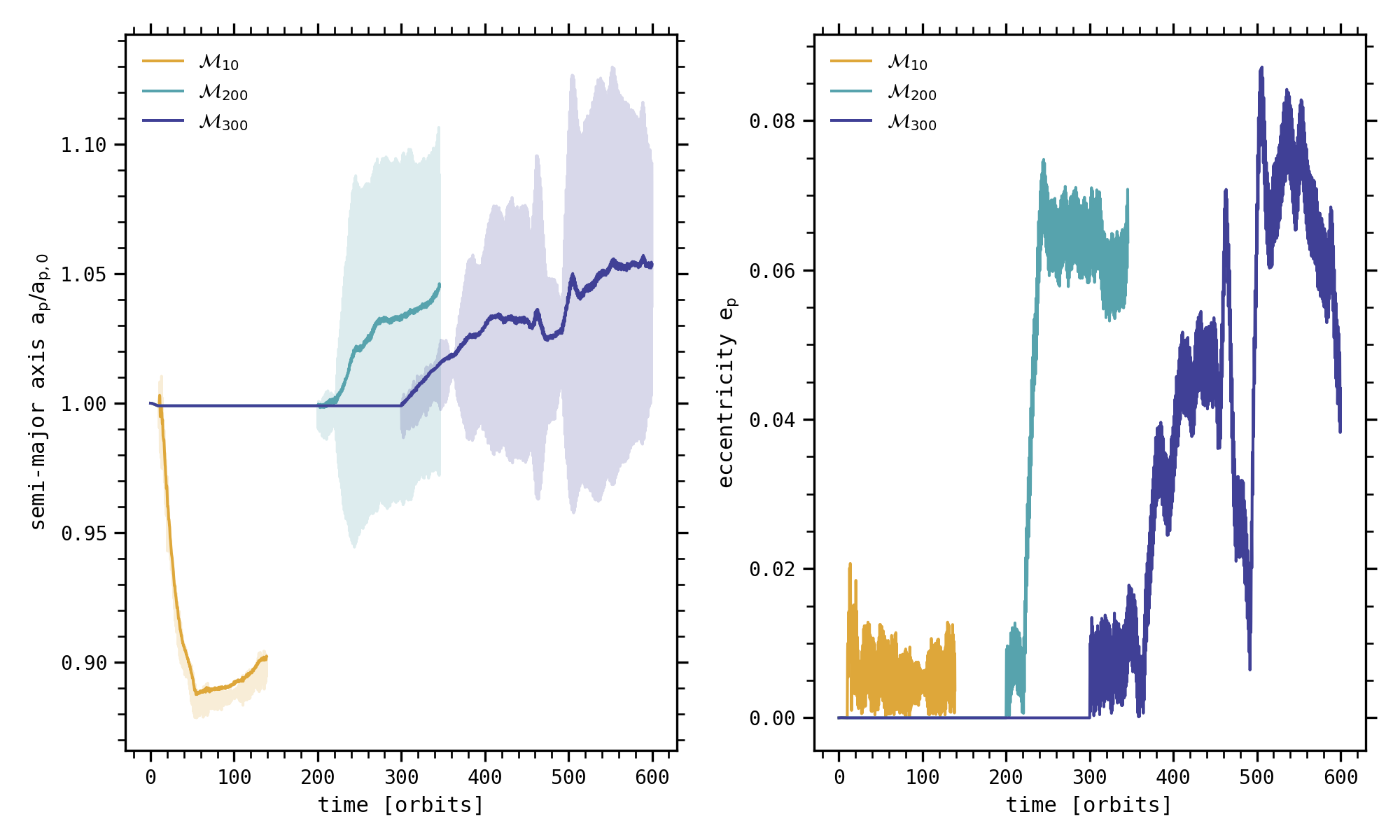}%
    \caption{Migration properties at different restart times and therefore for a different gap shape. Left: Temporal evolution of the semi-major axis $a_{\rm{p}}$ (solid lines) and radial position (shaded areas) of the planet. Whatever the initial condition, the planet migration is eventually slow and outward for $\beta_0=10^3$ once the gap has been carved. Right: Temporal evolution of the planet eccentricity $e_{\rm{p}}$, initially small ($<1.5\%$) before increasing abruptly ($\simeq7\%$) for the cases $\mathcal{M}_{200}$ and $\mathcal{M}_{300}$.}
    \label{fig:main_migration_eccentricity}
\end{figure*}

\subsection{Disk structures and variability}
\label{sec:disk_structures_variability}

We plot in Fig.~\ref{fig:main_spacetime} the space-time diagram of the azimuthally-averaged gas surface density $\Sigma$ for the $3$ simulations $\mathcal{M}_{300}$ (top panel), $\mathcal{M}_{200}$ (middle panel) and $\mathcal{M}_{10}$ (bottom panel). We first notice the presence of radial structures everywhere in the disk, with a planet-induced gap around $r=1$, but also planet-free gaps with locations that evolve with time (e.g., near $r\simeq2.7$ at $t=600$ orbits in the top panel). These planet-free structures are reminiscent of wind-induced gaps \citep{Bethune2017,Suriano2017,Suriano2018,Suriano2019,Riols2020,Cui&Bai2021}. We also see a strong variability inside the gap, with filaments of matter that disturb the planet gap, visible in particular in the top panel of Fig.~\ref{fig:main_spacetime} between $400$ and $600$ orbits, which suggests episodes of sporadic accretion. 

Most importantly, we retrieve the outward drift of the planet's outer gap, implying that the planet is gradually off-center with respect to its gap. When the planet is released, each simulation thus corresponds to a different gap shape, with no clear gap for $\mathcal{M}_{10}$, an episode of gas depletion with a quasi-symmetric gap for $\mathcal{M}_{200}$, and a strongly asymmetric gap for $\mathcal{M}_{300}$. 

In order to characterize planetary migration, we separate in each panel of Fig.~\ref{fig:main_spacetime} the fixed regime $t<T_\mathrm{release}$ from the migrating regime $t>T_\mathrm{release}$ by a vertical black dashed line. We also add an estimate of the gap edges (see Section~\ref{sec:gravitational_torques}) in black dots and the planet position with semi-transparent white lines. 

The planet being initially fixed at $r_{\rm{p,0}}=1$, the white line is initially horizontal at this location in all three panels. When $t>T_\mathrm{release}$, the planet in $\mathcal{M}_{300}$ and $\mathcal{M}_{200}$ does not seem to migrate, despite the massive disk that should impart some of its angular momentum to the planet. In addition, the planet eccentricity $e_{\rm{p}}$ seems to increase for these two runs after a few tens of orbits, such that the planet oscillates across a significant fraction of the gap's width. For $\mathcal{M}_{10}$ and after $10$ orbits, the planet has not carved a gap and seems to first migrate inwards until a deeper gap is formed (around $t=50$ orbits) reversing migration. The planet eccentricity remains low in this case.

\subsection{Planet migration and eccentricity}
\label{sec:planet_migration_eccentricity}

In order to better characterize how the planet migration is impacted by the planet gap shape, we plot in Fig.~\ref{fig:main_migration_eccentricity} the temporal evolution of the semi-major axis $a_{\rm{p}}$ (left panel) and the eccentricity $e_{\rm{p}}$ (right panel) of the planet for our 3 simulations.

$\mathcal{M}_{200}$ and $\mathcal{M}_{300}$ exhibit a semi-major axis that slowly increases with time, which implies a slow outward migration on average. For a planet initially at $10$ au, the migration rate would be on average $\simeq60$ au.Myr$^{-1}$ for $\mathcal{M}_{300}$ and $\simeq100$ au.Myr$^{-1}$ for $\mathcal{M}_{200}$. In these two cases, the right panel of Fig.~\ref{fig:main_migration_eccentricity} shows two main phases for the eccentricity: a short phase of low eccentricity ($<1\%$) for a few tens of orbits followed by an abrupt increase of the eccentricity ($\simeq6-8\%$). In $\mathcal{M}_{300}$ the eccentricity is globally increasing but quite variable, with an episode of lower eccentricity ($\simeq1-2\%$) near $500$ orbits corresponding to a migration stop in the left panel. 

In $\mathcal{M}_{10}$, the migration is activated as soon as the planet has reached its final mass at $10$ orbits and before the planet had carved a gap. Fig.~\ref{fig:main_migration_eccentricity} confirms that the migration is divided in two regimes: a relatively fast inward migration at a rate $\simeq730$ au.Myr$^{-1}$ for a planet initially at $10$ au followed by a slow outward migration at a rate $\simeq60$ au.Myr$^{-1}$ after an abrupt migration reversal at $\rm{t}=50$ orbits. In $\mathcal{M}_{10}$, the planet orbit remains quasi-circular during the whole migration. We checked that the planet inclination is negligible compared to the planet eccentricity by at least $2$ orders of magnitude.

\begin{figure}[htbp]
    \centering
    \includegraphics[width=0.99\hsize]{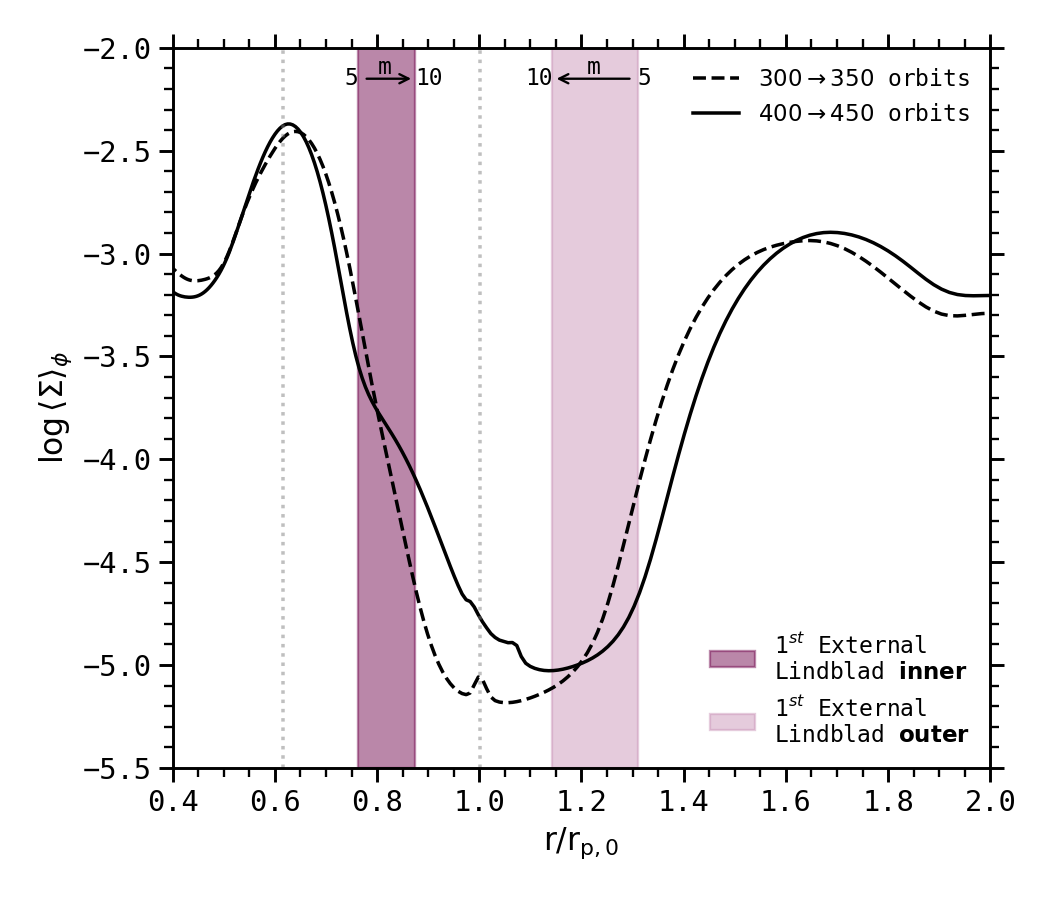}
    \caption{Azimuthal average of the logarithm of the gas surface density $\Sigma$ for the run $\mathcal{M}_{300}$, temporally averaged during a phase of quasi-circular and outward migration (dashed line), and during a phase of stalled and eccentric migration (solid line). The two vertical gray dotted lines show the reconstructed radii where the main sources of high-frequency $\Gamma_{\rm{Li}}$ variability are expected to originate from. The purple areas indicate the location of the inner and outer first-order external Lindblad resonances, for the modes $\rm{m}=5$ to $\rm{m}=10$. For the quasi-circular case (dashed line), the wind has already affected the planet gap, with a wider outer gap compared to the inner gap. For the eccentric case (solid line), the gap is globally denser due to the increase of the planet eccentricity. The wind has strongly enhanced the inner gap compared to the outer gap that is even further away from the planet location. The negative density gradient at the planet location could boost a negative corotation torque.}
    \label{fig:main_axisigma}
\end{figure}

We show a radial profile of the surface density $\Sigma$ in $\mathcal{M}_{300}$ in Fig.~\ref{fig:main_axisigma}, overlaid in purple with the locations of first-order external Lindblad resonances for $m=5$ to $m=10$. $\Sigma$ is here averaged azimuthally and temporally over $50$ orbits for two epochs: when the planet migrates outward and is nearly circular at $t\in[300-350]$ orbits ($e_{\rm{p}}\lesssim1.5\%$, black dashed line) and once the planet stalls and its eccentricity is close to $5\%$ at $t\in[400-450]$ orbits (black solid line). At early times, we already detect a clear asymmetry of the gap shape that should participate in the torque imbalance and the outward planet migration, because outer principal resonances are expected to fall in the wide and deep outer gap, while inner principal resonances are expected to be enhanced by material accumulating close to the narrower and shallower inner gap.

By comparing the gap shape between both epochs, we first notice that there is a clear shift outwards of the solid line compared to the dashed line, at the location of the outer density maximum. This drift illustrates the progressive erosion of the outer disk by the gap and is due to the fast accretion driven by the wind in the gap \citep[][\citetalias{Wafflard-Fernandez&Lesur2023}]{Martel&Lesur2022}. As a result, mass progressively piles up in the inner gap, leading to the strong gap dissymmetry observed at later times. This negative density gradient locally around the planet is expected to negatively reinforce the corotation torque exerted onto the planet \citep{Peplinski2008a,Peplinski2008b}, slowing down its migration.

Second, the gap is globally shallower at later times (solid line). This could be due to the increased planet eccentricity since one expects a shallower gap around a massive eccentric planet compared to a circular one \citep[see Fig.~4 in][]{Duffell&Chiang2015}. 


\subsection{Gravitational torques}
\label{sec:gravitational_torques}

\begin{figure*}
    \centering
    \includegraphics[width=0.955\hsize]{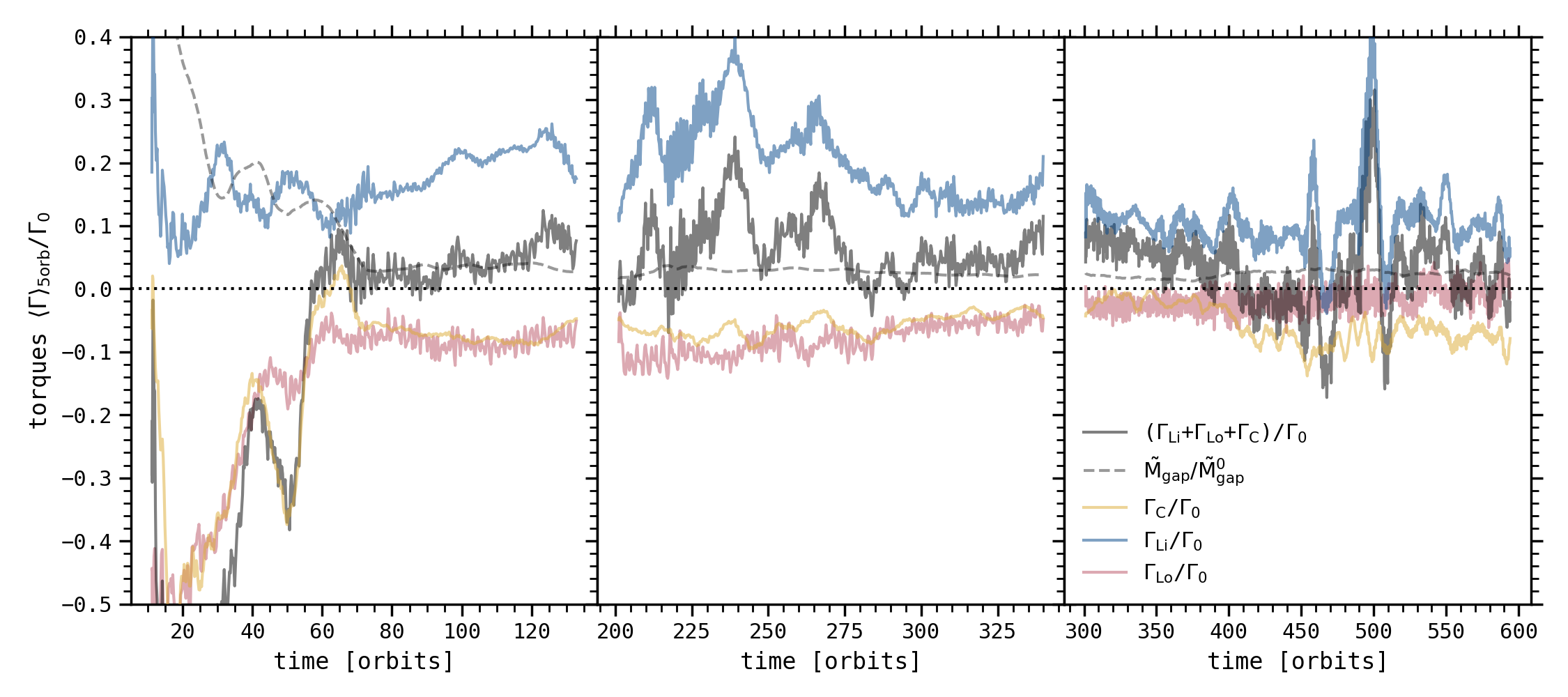}\
    \caption{Temporal evolution of various torque components exerted by the gas on the migrating planet and displayed with a moving average of $5$ orbits, for the three runs $\mathcal{M}_{10}$ (left panel), $\mathcal{M}_{200}$ (middle panel) and $\mathcal{M}_{300}$ (right panel). The total torque (black solid curves) is decomposed in $3$ components, coming from different regions: the torque from the inner disk associated to the inner Linblad torque $\langle\Gamma_{\rm{Li}}\rangle_{\rm{5orb}}$ (blue curves, $\langle\Gamma_{\rm{Li}}\rangle_{\rm{5orb}}\gtrsim0$), the torque from the outer disk associated to the outer Linblad torque $\langle\Gamma_{\rm{Lo}}\rangle_{\rm{5orb}}$ (red curves, $\langle\Gamma_{\rm{Lo}}\rangle_{\rm{5orb}}\lesssim0$) and the torque in the planet gap associated to the corotation torque $\langle\Gamma_{\rm{C}}\rangle_{\rm{5orb}}$ (yellow curves, $\langle\Gamma_{\rm{C}}\rangle_{\rm{5orb}}\lesssim0$). All these torques are normalized by $\Gamma_0$. A proxy for the mass reservoir enclosed in the gap is represented via the black dashed line.}
    \label{fig:main_torque}
\end{figure*}

In all cases, the planet eventually migrates outward once it has carved a deep gap, which seems quite robust for such a high value of the initial magnetization. In order to explain this migration behavior, we can study the gravitational torques\footnote{In general we cannot interpret the direction and amplitude of migration of an eccentric planet directly from the temporal evolution of the gravitational torque only, and we need to rather rely on the gravitational power. We checked here that the sign and qualitative behavior of the power and the torque are close to one each other, except for a slightly stronger variability for the power.
} of the gas onto the planet.

We plot in Fig.~\ref{fig:main_torque} the temporal evolution of the dimensionless gravitational torque $\langle\Gamma\rangle_{\rm{5orb}}/\Gamma_0$ exerted by the disk onto the planet with a moving average of $5$ orbits and $\Gamma_0=\displaystyle\left(q_p/h_0\right)^2\Sigma_0 r_p^4 \Omega^2_p$, evaluated initially and at $r_{\rm{p,0}}$. We divide $\Gamma$ in three components: the corotation torque $\Gamma_{\rm{C}}$ in yellow, the inner Lindblad torque $\Gamma_{\rm{Li}}$ in blue and the outer Lindblad torque $\Gamma_{\rm{Lo}}$ in red. We define the corotation region as the region where the fluid azimuthally-averaged horizontal velocity in the planet's frame deviates significantly from a purely Keplerian flow. We choose the criterion $|\langle v_{\rm{r}} \rangle_{\phi}|/|\langle v_{\rm{\phi}} \rangle_\phi|\gtrsim4\% $ in the midplane, smoothing the velocity fields radially and temporally over $1$ orbit so as to capture almost the entire planet gap without being too sensitive to the local variability of the velocity fields. As mentioned in Section~\ref{sec:disk_structures_variability}, the resulting estimate of the planet gap width is represented with the black lines in Fig.~\ref{fig:main_spacetime}. We then compute the gravitational torque exerted by the gas inside the corotation region $\langle\Gamma_{\rm{C}}\rangle_{\rm{5orb}}$ by summing the contribution of all the cells inside that region, excluding the effect of the gas inside the Hill sphere. Note that it is slightly different from what is computed in the simulations, as part of this material participates to the total torque, given a $\sin^2(x)$ tapering function between $50\%$ and $100\%$ of the Hill sphere. The Lindblad torques are simply the sum of the contribution of the cells inside the planet gap's inner edge ($\langle\Gamma_{\rm{Li}}\rangle_{\rm{5orb}}$) and outside the planet gap's outer edge ($\langle\Gamma_{\rm{Lo}}\rangle_{\rm{5orb}}$). The black dashed lines give an estimate of the gas mass included inside the corotation region, removing the contribution from the Hill sphere, and normalizing by a proxy of the mass reservoir initially available inside the gap. We therefore expect this quantity to decrease as a gap is carved by the planet, before reaching a quasi-constant value. Looking at the temporal evolution of these various gravitational torques, we see that the inner Lindblad torque is globally positive for all cases, whereas the corotation torque and the outer Lindblad torque tend to be negative.

\begin{enumerate}
    \item In $\mathcal{M}_{10}$, the corotation torque and the outer Lindblad torque are strongly negative and dominate during the first $50$ orbits, as a significant mass of gas material is contained inside the corotation region. Beyond $50$ orbits, the corotation region has been depleted and reaches a few percents of the initial mass reservoir (see black dashed line), which decreases the amplitude of the corotation torque, whereas the inner Linblad torque which slowly increases is eventually multiplied by $2$, and becomes the main driver of migration. This behavior of the torques confirms the migration pattern for this simulation, with a first phase of type-III-like inward migration with a partial gap, followed by a sudden migration stop when the dynamical corotation torque cancels out, and a migration reversal dominated by the behavior of the gas material inside the inner gap edge \citep[see also Fig.~6 in][]{Hu2025}. \\
    \item For $\mathcal{M}_{200}$, $|\langle\Gamma_{\rm{Lo}}\rangle_{\rm{5orb}}|$  is quite variable but globally decreases. This is probably because the outer gap erosion is faster than the outward planet migration rate, which decreases the contribution of the gas material outside the gap's outer edge to $\langle\Gamma_{\rm{Lo}}\rangle_{\rm{5orb}}$. For the same period of time, $|\langle\Gamma_{\rm{C}}\rangle_{\rm{5orb}}|$ is globally divided by $\sim2$ as the mass inside the gap is also divided by $\sim2$. The main driver of migration is therefore $\langle\Gamma_{\rm{Li}}\rangle_{\rm{5orb}}>0$, with episodes of strong amplification of $\langle\Gamma_{\rm{Li}}\rangle_{\rm{5orb}}$ corresponding to phases of faster outward migration. \\
    \item The behavior of the torques is roughly similar for the run $\mathcal{M}_{300}$. The total torque is globally positive when $t<380$ orbits, with a contribution of the torque components similar to that of $\mathcal{M}_{200}$. However, the behavior of the inner and corotation torques $\langle\Gamma_{\rm{Li}}\rangle_{\rm{5orb}}$ and $\langle\Gamma_{\rm{C}}\rangle_{\rm{5orb}}$ departs from what has been described earlier after $t\simeq380$ orbits. In particular, there is a stronger variability inside the gap, associated with an increase of both the mass inside the gap and $|\langle\Gamma_{\rm{C}}\rangle_{\rm{5orb}}|$ by a factor $\sim2$. We observe also that the gap's inner edge slightly shrinks and gets closer to the planet location (black line in the top panel of Fig.~\ref{fig:main_spacetime}), as the planet eccentricity exceeds $\simeq2\%$. The end result is that $\langle\Gamma_{\rm{Li}}\rangle_{\rm{5orb}}$ is on average counterbalanced by $\langle\Gamma_{\rm{C}}\rangle_{\rm{5orb}}$, leading to a total torque that is nearly zero on average at $t\simeq400$ orbits, and therefore a migration stall. In addition,  the strong variability of $\langle\Gamma_{\rm{Li}}\rangle_{\rm{5orb}}$ is sufficient to induce strong fluctuations in the semi-major axis and eccentricity of the planet (e.g., near $t=460,500$ orbits). \\
\end{enumerate}

This discussion highlights the correlation between the variability of $\langle\Gamma_{\rm{Li}}\rangle$ and the planet stochastic behavior. While a part of this variability might be associated with sporadic episodes of accretion through the gap driven by the wind torque, we find that $\langle\Gamma_{\rm{Li}}\rangle$ is also subject to rapid quasi-periodic oscillations at two characteristic frequencies: $\omega=\omega_p$ and  $\omega\simeq 2\omega_p\equiv \omega_c$ (see Appendix~\ref{sec:appendix_fft}). While the first frequency can easily been understood as resulting from the planet's eccentricity, the second one is more elusive. We can relate these frequencies to two corotation radii that we represent as two vertical gray dotted lines in Fig.~\ref{fig:main_axisigma}. We find that $\omega_c$ corresponds to the peak of the inner density ring lying at the gap edge around $r\simeq 0.63r_{\rm{p,0}}$ (see also top panel of Fig.~\ref{fig:main_spacetime}). Density rings are known to be subject to the Rossby-Wave Instability \citep[RWI,][]{Lovelace1999}, and we observe that an elongated vortex indeed develops at this location (see left panel of Fig.~\ref{fig:appendix_horseshoe}). Therefore, we associate the quasi periodic oscillation at $\omega_c$ of the inner Lindblad torque to a transient vortex triggered at the inner gap edge, lasting a few tens of orbits, but regenerated sporadically.

The corotation torque, and more specifically the gravitational torque measured in the gap region, is negative for two reasons. First, the contribution of the gas material accumulated near the outer wake and episodically carried inwards by the accretion flow accounts for a significant fraction of the total $\langle\Gamma_{\rm{C}}\rangle_{\rm{5orb}}$ (see Appendix~\ref{sec:appendix_horseshoe} for the shape of the horseshoe region). Second, the gap radial asymmetry leads to a negative density gradient at the planet location. This gradient is sharper at later times for the run $\mathcal{M}_{300}$ (see solid line in Fig.~\ref{fig:main_axisigma}) and could be responsible for the enhanced $\langle\Gamma_{\rm{C}}\rangle_{\rm{5orb}}$ at $t>400$ orbits in the right panel of Fig.~\ref{fig:main_torque}.

\subsection{Origin of the eccentricity}
\label{sec:main_eccentricity}

The aim of this section is to have a qualitative understanding of the processes behind the evolution of the planet eccentricity. To that end, we focus on the epoch of $e_{\rm{p}}$ amplification for the run $\mathcal{M}_{200}$ at $t\in[206-248]$ orbits (see turquoise curve in the right panel of Fig.~\ref{fig:main_migration_eccentricity}), and follow the semi-analytical work from \cite{Duffell&Chiang2015}. In particular, we make extensive use of their Appendix~A and the formulae needed to evaluate the contributions to $\dot{e}_{\rm{p}}$ from various resonances \citep{Goldreich&Tremaine1980,Papaloizou&Larwood2000,Ogilvie&Lubow2003}. We focus on the principal Lindblad resonances (\labelization{LR0}), the first-order co-orbital (\labelization{LR1c}) and external (\labelization{LR1e}) Lindblad resonances, and the first-order corotation resonances (\labelization{CR1}). These resonances are called first-order resonances as the perturbing potential of an eccentric planet can be expanded as a principal component and a first-order component in the eccentricity \citep{Masset2008}. Note that we do not write here the full set of equations required to compute $\dot{e}_{\rm{p}}$ for all these resonances but refer the reader to the Appendix~A of \cite{Duffell&Chiang2015} for more details. 

A general formula to derive the contribution of a first-order resonance $\mathcal{R}$ to $\dot{e}_{\rm{p}}$ can be written as a sum over azimuthal wave numbers $m$: 
\begin{equation}
    \dot{e}_{\rm{p}}(\mathcal{R}) = \sum_{m=1}^{m_\mathrm{max}} \pm \frac{T_\mathcal{R}(m)}{m L_{\rm{p}}} \sqrt{\frac{1-e_{\rm p}^2}{e_{\rm{p}}^2}} \mathcal{F}_w(\mathcal{R}),
    \label{eq:depl}
\end{equation}

\noindent where $\displaystyle L_{\rm{p}}=q_{\rm{p}}\sqrt{GM^3_\star a_{\rm{p}}(1-e_{\rm{p}}^2)}$ is the planet's angular momentum and depends on the planet properties ($a_{\rm{p}}$, $e_{\rm{p}}$, $q_{\rm{p}}$), $m_{\rm{max}}=(2h)^{-1}$ is a rough approximation of the torque cut-off for a non-self-gravitating disk \citep{Goldreich&Tremaine1980, Duffell&Chiang2015}. $\mathcal{F}_w(\mathcal{R})$ is a weakening function that may decrease the contribution to $\dot{e}_{\rm{p}}$ of the $m$-fold component of the resonance-dependent torque $T_\mathcal{R}(m)$, whether it be because of the increasingly supersonic motion of the epicyclic frequency for the Lindblad torques \citep{Papaloizou&Larwood2000,Fairbairn&Rafikov2025} or a saturation function for the corotation torque \citep{Ogilvie&Lubow2003,Goldreich&Sari2003}. 

The methodology that we choose here is semi-analytic as it relies on physical quantities estimated in $\mathcal{M}_{200}$ at the location of the resonances, using the Keplerian frequency $\Omega_{\rm{p}}$ at the planet location. Thus, we measure on the one hand $\Sigma(r)$ at the location of the Lindblad resonances (\labelization{LR0}, \labelization{LR1c} and \labelization{LR1e}), and on the other hand we need $\nu(r)$, $\Sigma(r)$ and $d\Sigma(r)/dr$ for \labelization{CR1}. For all these quantities, we systematically remove the contribution of the material inside the Hill sphere, as this region contains large and highly variable over-densities that analytic perturbative theories do not intend to treat. Because the turbulent viscosity in our simulations is not parametrized via an $\alpha_\nu$ parameter \citep{Shakura1973}, we first measure for each snapshot in the simulation the radial Reynolds stress, and from this estimate we choose a constant $\nu=1.25\times10^{-6}$, corresponding to $\alpha_\nu=5\times 10^{-4}$ at $r=r_{\rm{p,0}}$. The aim is mainly to provide a simplified illustration of the behavior of the first-order corotation resonances, especially since we use here a viscous formalism that is expected to be far from the real conditions of our simulations.

\begin{figure}[htbp]
    \centering
    \includegraphics[width=0.99\hsize]{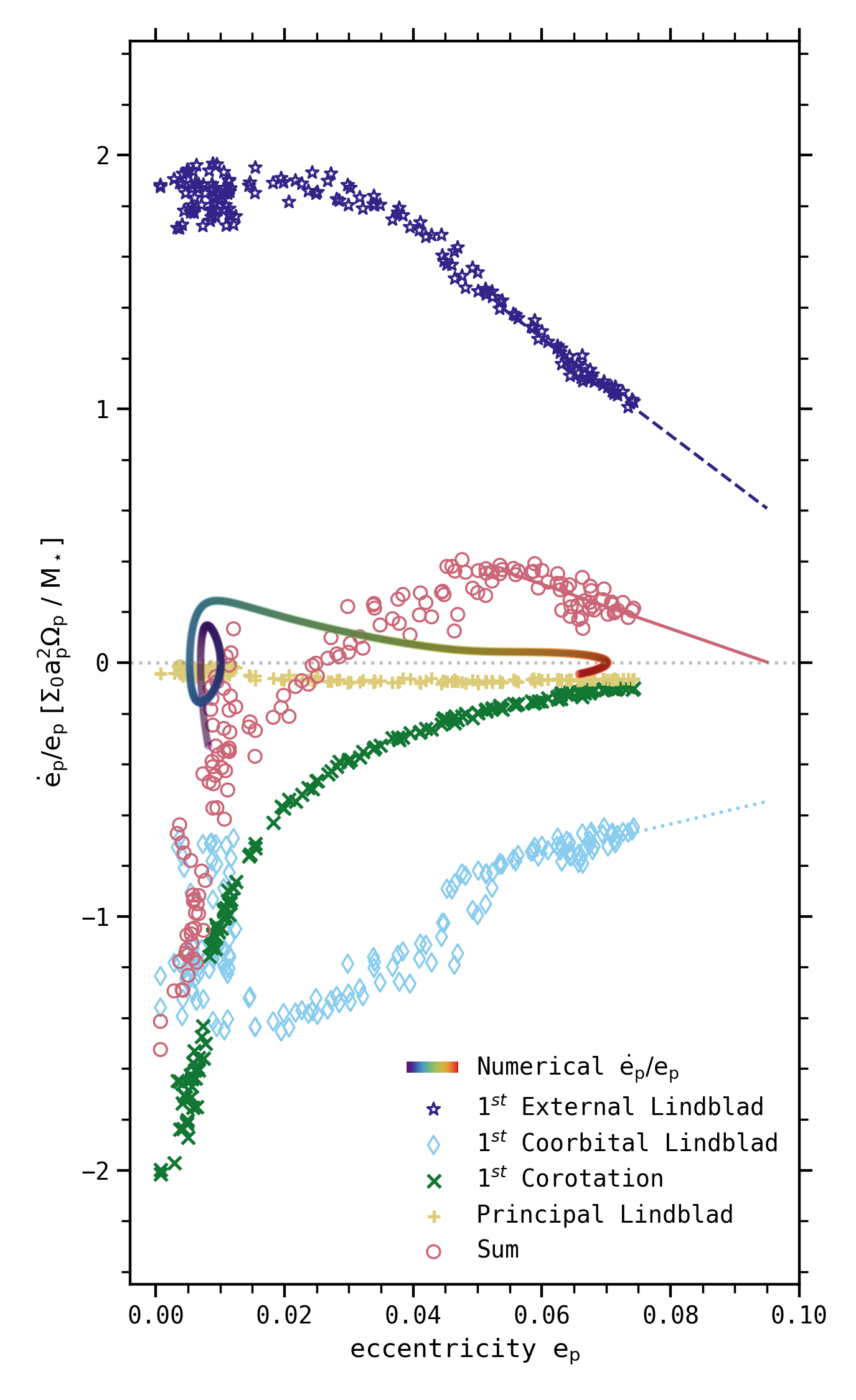}%
    \caption{Contribution of various resonances to the eccentricity evolution $\dot{e}_{\rm{p}}/e_{\rm{p}}$ as a function of $e_{\rm{p}}$, for the run $\mathcal{M}_{200}$ and during the episode of $e_{\rm{p}}$ amplification at $t\in[206-248]$ orbits. The semi-analytical decomposition focuses on the principal Lindblad resonances (\labelization{LR0}, yellow $+$), the first-order corotation resonances (\labelization{CR1}, green $\times$), the first-order co-orbital (\labelization{LR1c}, light blue $\lozenge$) and external (\labelization{LR1e}, dark blue $\star$) Lindblad resonances. The sum of all these contributions are represented with red $\circ$. The rainbow curve represents a numerical measure of $\dot{e}_{\rm{p}}/e_{\rm{p}}$ between $t=206$ orbits (purple color) and $t=248$ orbits (red color). Eccentricity growth is a finite-amplitude instability, with a damping of $e_{\rm{p}}$ when $e_{\rm{p}}<1\%$ or $e_{\rm{p}}\gtrsim7\%$, and an excitation of $e_{\rm{p}}$ when $e_{\rm{p}}\in[1.5\%-7\%]$.}
    \label{fig:main_depl}
\end{figure}

We plot in Fig.~\ref{fig:main_depl} the various contribution to the eccentricity evolution $\dot{e}_{\rm{p}}/e_{\rm{p}}$ as a function of $e_{\rm{p}}$, for the run $\mathcal{M}_{200}$ at $t\in[206-248]$ orbits. Because $e_{\rm{p}}$ is roughly increasing with time during our epoch of interest, a higher $e_{\rm{p}}$ in this graph indicates also a later period in the simulation. For a given planet eccentricity, $\dot{e}_{\rm{p}}/e_{\rm{p}}<0$ means that the eccentricity will be damped, and the planet will tend to be stabilized on a circular orbit. Conversely, $\dot{e}_{\rm{p}}/e_{\rm{p}}>0$ means that the eccentricity will be excited, and the planet will tend to be pushed into an even more eccentric orbit.

We first focus in Fig.~\ref{fig:main_depl} on the rainbow curve, which represents a smoothed numerical measure of $\dot{e}_{\rm{p}}/e_{\rm{p}}$ as a function of $e_{\rm{p}}$, the colors indicating the temporal evolution from purple ($t=206$ orbits) to red ($t=248$ orbits). When $e_{\rm{p}}\lesssim1\%$ (purple to blue), $\dot{e}_{\rm{p}}/e_{\rm{p}}$ alternates between negative and positive values, indicative of oscillations between phases of $e_{\rm{p}}$ damping and $e_{\rm{p}}$ driving. If $e_{\rm{p}}$ somehow reaches higher values $e_{\rm{p}}\in[1\%-7\%]$ (green to orange), $\dot{e}_{\rm{p}}/e_{\rm{p}}$ is always positive, and $e_{\rm{p}}$ increases even more. Finally, when $e_{\rm{p}}\gtrsim7\%$ (red), $\dot{e}_{\rm{p}}/e_{\rm{p}}$ switches from positive to slightly negative values, meaning that $e_{\rm{p}}$ reaches a plateau and slowly decreases. This picture is consistent with the right panel of Fig.~\ref{fig:main_migration_eccentricity} and is compatible with a finite-amplitude instability scenario \citep{Goldreich&Sari2003}, which would be activated when $e^{\rm{min}}_{\rm{p}}=1\%-1.5\%$, and would be limited by the increasingly supersonic motion of the planet's epicyclic frequency $\kappa_{\rm{p}}$. 

\begin{figure*}
    \centering
    \includegraphics[width=0.99\hsize]{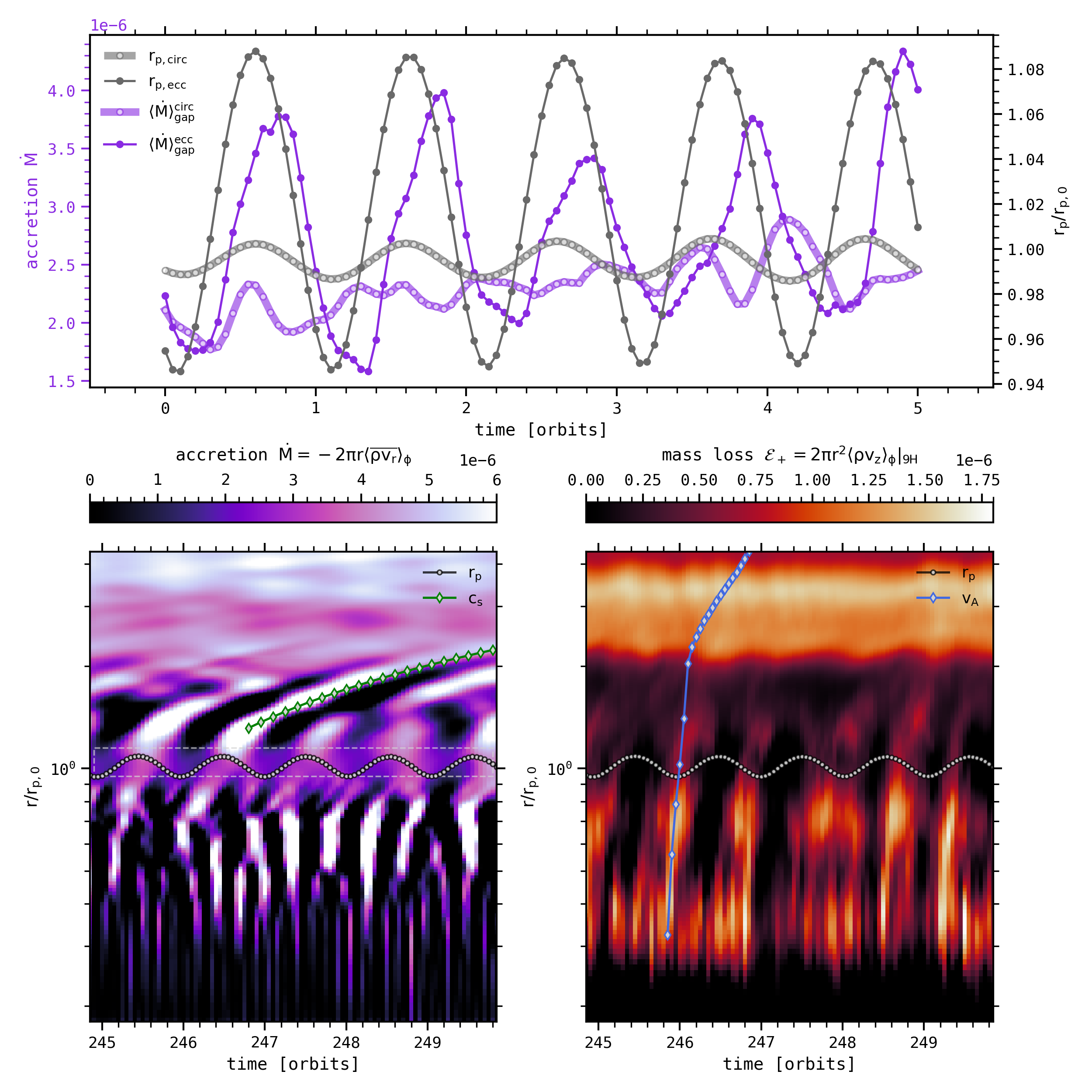}\
    \caption{Eccentricity-driven modulation of the accretion and ejection processes. Top: $5$-orbit evolution of the planet radius $r_{\rm{p}}$ (gray lines) and accretion rate in the planet gap $\langle\dot{M}\rangle_{\rm{gap}}$ (purple lines) during two epochs in the run $\mathcal{M}_{200}$: when the planet orbit is quasi-circular (thick lines) and once the planet eccentricity is nearly maximum (thin lines). Bottom: space-time diagram of the accretion rate $\dot{M}$ (left) and the ejection $\mathcal{E}_+$ at $\arctan(9h)$ (right) during the $5$-orbit evolution of the eccentric giant planet. The instantaneous radial position of the planet is shown with white circles. Some characteristic velocities indicate how planet-driven waves are propagating in the disk: the adiabatic sound speed $c_s$ in green, and the Alfvén speed $v_A$ in blue. The gap location used to compute $\langle\dot{M}\rangle^{\rm{ecc}}_{\rm{gap}}$ in the top panel is indicated by a white dashed rectangle in the bottom left panel.}
    \label{fig:main_modulation}
\end{figure*}

In order to test that scenario, we plot in Fig.~\ref{fig:main_depl} contributions to the eccentricity evolution $\dot{e}_{\rm{p}}/e_{\rm{p}}$: \labelization{LR0}, \labelization{CR1}, \labelization{LR1c} and \labelization{LR1e} as a function of $e_{\rm{p}}$. The sum of these contributions is represented by the red circles, to be compared qualitatively with the rainbow curve. We show with these red circles that $\dot{e}_{\rm{p}}/e_{\rm{p}}$ is negative whenever $e_{\rm{p}}\ll2\%$, is positive and increasing when $e_{\rm{p}}\geq2\%$, and eventually decreases when $e_{\rm{p}}/h_0\gtrsim1$. The global evolution of $\dot{e}_{\rm{p}}/e_{\rm{p}}$ is therefore consistent with a finite-amplitude instability: (i) $e_{\rm{p}}$ is damped as long as the planet orbit is nearly circular. If the eccentricity reaches up to $\simeq1.5\%$, the planet is pushed back to a quasi-circular orbit. This behavior can be seen for all simulations in the right panel of Fig.~\ref{fig:main_migration_eccentricity}, where the planet remains confined in the low-eccentricity regime and oscillates between $0$ and $1.5\%$ over a significant period of time (about $140$, $20$ and $75$ orbits for $\mathcal{M}_{10}$, $\mathcal{M}_{200}$ and $\mathcal{M}_{300}$ respectively). (ii) Beyond $e_{\rm{p}}\simeq2\%$, the eccentricity is amplified until $e_{\rm{p}}\simeq h_0$, beyond which $\dot{e}_{\rm{p}}/e_{\rm{p}}$ is still positive but the slope is negative, meaning that the eccentricity is less and less amplified. To understand this mechanism of saturation when $e_{\rm{p}}/h_0\simeq1$, we define the velocity $v_\epsilon$ of the epicyclic motion as the product of half the epicyclic frequency $\kappa_{\rm{p}}$ ($=\Omega_{\rm{p}}$ for a Keplerian orbit) and the radial distance the planet crosses between its apastron radius $r_\alpha=(1+e_{\rm{p}})a_{\rm{p}}$ and its periastron radius $r_\pi=(1-e_{\rm{p}})a_{\rm{p}}$ :
\begin{equation}
    v_\epsilon = \displaystyle\frac{\kappa_{\rm{p}}}{2}(r_\alpha-r_\pi) = \frac{e_{\rm{p}}}{h_0}c_{s,0},
    \label{eq:supersonic}
\end{equation}

\noindent using the isothermal sound speed $c_{s,0}$. $v_\epsilon$ is therefore supersonic when $e_{\rm{p}}/h_0>1$, which may explain why the $e_{\rm{p}}$ eventually stabilizes as it approaches $6-8\%$ \citep{Fairbairn&Rafikov2025}.

Looking at the contributions of the various resonances, we show that principal Lindblad resonances  tend to damp the planet eccentricity \citep[$\dot{e}_{\rm{p}}(\mathcal{R}=\textrm{\labelization{LR0}})<0$, consistent with an outward migration, see, e.g.,][their Section.~5]{Masset2008} but are negligible compared to first-order resonances, in agreement with \cite{Goldreich&Sari2003,Duffell&Chiang2015}. The first-order external Lindblad resonances \labelization{LR1e} are the only source of $e_{\rm{p}}$ excitation and dominates the behavior of $\dot{e}_{\rm{p}}$ when $e_{\rm{p}}>2\%$. 
They are counterbalanced at low $e_{\rm{p}}\ll2\%$, by the ﬁrst-order co-orbital Lindblad \labelization{LR1c} and corotation resonances (\labelization{CR0}). However, these two damping mechanisms vanish as the gap density decreases (\labelization{LR1c}) and the saturation function $\mathcal{F}_w(\mathcal{R})$ decreases  when increasing $e_{\rm{p}}$ (\labelization{CR0}). It is important to stress that our semi-analytic methodology relies on the estimate of $\nu$, which strongly impacts $\mathcal{F}_w(\mathcal{R})$, and therefore determines precisely how fast the amplitude of $\dot{e}_{\rm{p}}(\mathcal{R}=\textrm{\labelization{CR1}})/e_{\rm{p}}<0$ goes to zero when increasing the eccentricity. Hence, our interpretation in only semi-quantitative.

\subsection{Eccentricity-driven accretion/ejection modulation}
\label{sec:main_modulation}

Now that we have a qualitative explanation of how the planet eccentricity is excited by the disk via first-order external Lindblad resonances, we seek to determine in this section the back-reaction of such a moderate eccentricity on the gaseous disk. 
To analyze the time-variability of the disk, we zoom on 2 epochs of $5$ orbits in $\mathcal{M}_{200}$: when the planet is quasi-circular at $t=206.4$ orbits (epoch $\mathcal{T}_{\rm{circ}}$), thus before the episode of $e_{\rm{p}}$ amplification, and when the planet reaches a nearly maximum eccentricity at $t=244.85$ orbits (epoch $\mathcal{T}_{\rm{ecc}}$).

\begin{figure*}
    \centering
    \includegraphics[width=0.99\hsize]{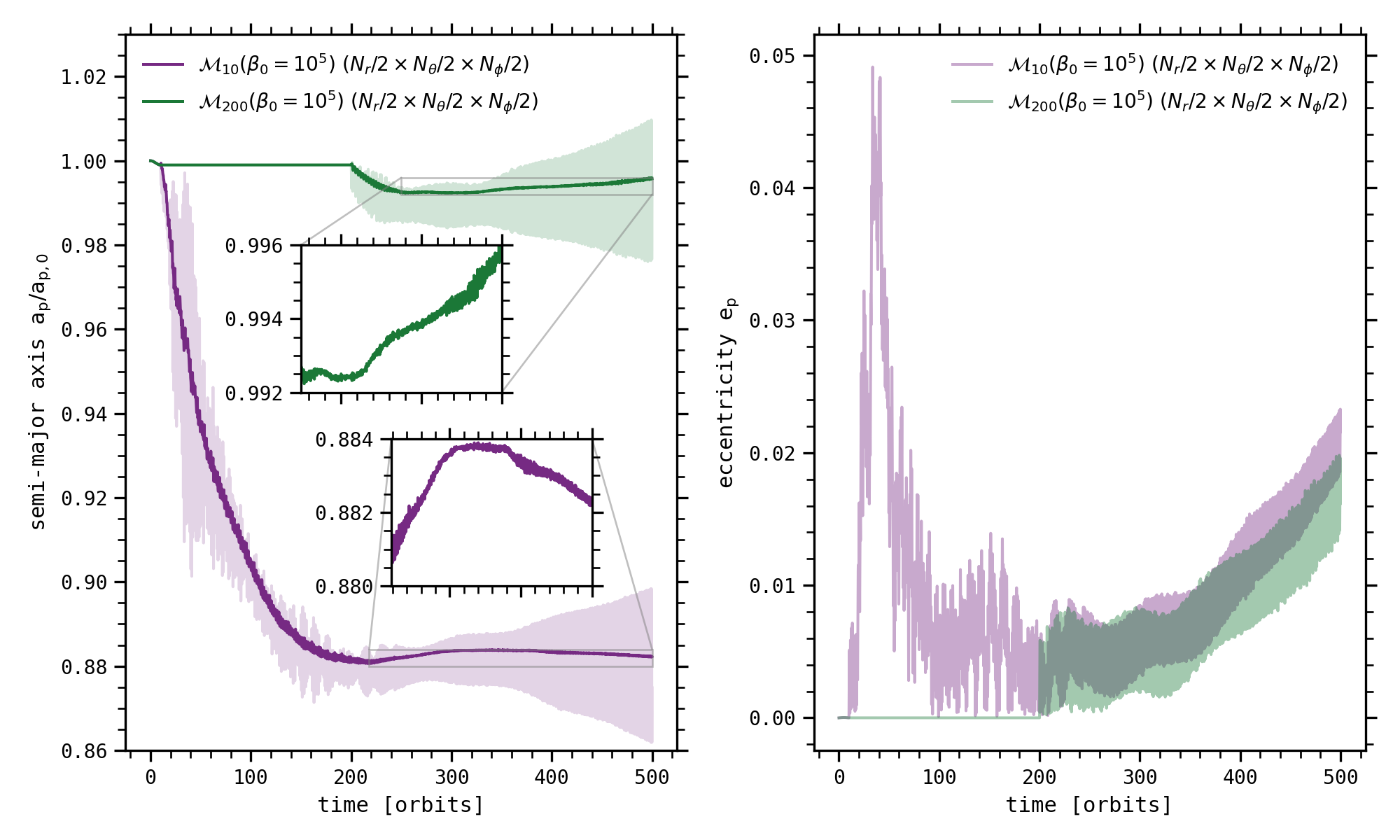}%
    \caption{Same as Fig.~\ref{fig:main_migration_eccentricity}, but for a lower magnetization $\beta_0=10^5$, and dividing the resolution by $2$ in every direction. Migration is activated after $10$ and $200$ planet orbits for the violet and green curves respectively. Once a deep gap has been carved, migration is slow and $e_{\rm{p}}$ slowly increases.}
    \label{fig:main_migration_eccentricity_b5}
\end{figure*}

The $5$-orbit evolution of the planet radius $r_{\rm{p}}$ is shown in the top panel of Fig.~\ref{fig:main_modulation} for both $\mathcal{T}_{\rm{circ}}$ (thick gray line) and $\mathcal{T}_{\rm{ecc}}$ (thin gray line). 
We also plot in the top panel of Fig.~\ref{fig:main_modulation} the $5$-orbit evolution of the accretion rate $\dot{M}=-2\pi r \langle \overline{\rho v_r} \rangle_\phi$ through the gap. Each dot represents one snapshot of the simulation from which we reconstruct the accretion behavior in the planet gap. 
Interestingly, $\dot{M}$ varies weakly in the circular case (thick purple line), but exhibits a clear periodic modulation in the eccentric case (thin purple line), with a ratio between the maximum and minimum reaching $2.5$. Because this modulation of $\dot{M}$ is not clearly detectable in the circular case and has a typical frequency close to $\kappa_{\rm{p}}$ in the eccentric case, we associate it to the planet eccentricity. 

In the rest of this section, we focus on the eccentric case. In the bottom left panel of Fig.~\ref{fig:main_modulation}, we plot the space-time diagram of $\dot{M}$. 
The eccentric orbit of the planet is represented by the white dots. We retrieve the eccentricity-driven modulation of the accretion, particularly visible in the outer gap and partly in the outer disk ($r/r_{\rm{p,0}}<3$). We notice also the propagation of sonic acoustic waves from the planet position towards the outer disk, due to the epicyclic motion of the planet. The characteristic of these acoustic waves are represented in the bottom left panel of Fig.~\ref{fig:main_modulation} with green lozenges and match the perturb $\dot{M}$ accurately.

In the bottom right panel of Fig.~\ref{fig:main_modulation}, we plot the space-time diagram of the wind mass lass rate $\mathcal{E}_+=2\pi r^2 \langle \overline{\rho v_z} \rangle_\phi|_{\rm{9H}}$, linked to the vertical mass flux and estimated at $\Theta_+ = \arctan(9h)$, well above the disk upper surface. $\mathcal{E}_+$ is negative at $\theta=\Theta_+$ if the flow locally falls back towards the disk midplane. We show that some material is periodically ejected vertically by the wind at $\Theta_+$ from $\simeq0.6r_{\rm{p,0}}$ to $\simeq2r_{\rm{p,0}}$, and especially in the inner part of this wind region. The period of appearance of these ejecta is consistent with the epicyclic frequency of the eccentric planet. In order to apprehend the typical speed of propagation of these ejecta, we represent with blue lozenges the characteristic on an alfvén wave in the bottom right panel of Fig.~\ref{fig:main_modulation}. The relatively good match between the modulation of the wind ejection and the Alfv\'enic characteristic indicate that the planet eccentricity launches a modulation in the disk magnetosphere that travels along the magnetic field lines, possibly on long distances in the outflow. 

We show in Fig.~\ref{fig:appendix_modulation} of Appendix~\ref{sec:appendix_modulation} that the modulation of the wind mass loss rate is top-down asymmetric, with no modulation of $\mathcal{E}_-$ from the more massive wind estimated at $\theta=\Theta_-=-\Theta_+$, well below the disk's lower surface. Due to the slight inclination of poloidal magnetic field lines at the disk surface, we expect the launching point of the modulated ejected material to originate from an upper accretion layer in the inner regions of the disk after an episode of strong $\dot{M}$, but not from a lower accretion layer. This picture is indeed consistent with the fact that the accretion flows are also top-down asymmetric in the inner regions, with material crossing the inner gap and being elevated to a single accretion layer at the upper surface. Although this $\mathcal{E}_+$ variability could be a physical process due to a significant fraction of accreted material being ejected from the inner disk after an episode of strong accretion, we cannot rule out the potential influence of the inner boundary conditions. Whatever the origin of its launching mechanism, the modulation of the mass loss rate is related to the forcing of the planet eccentricity.

\subsection{Impact of the disk magnetization}
\label{sec:magnetization}


The simulations presented above were all performed with a relatively strong large-scale field $\beta=10^3$, and a natural question is to check the impact of weaker field on the planet's dynamics. We have therefore performed preliminary simulations with a lower magnetization ($\beta_0=10^5\equiv\beta_5$) and a lower resolution ($N_r/2 \times N_\theta/2 \times N_\phi/2 = 256 \times 128 \times 512$). We show in Fig.~\ref{fig:main_migration_eccentricity_b5} the evolution of the semi-major axis (left panel) and the eccentricity (right panel) of the Jovian planet, activating the planet migration after the planet has reached its final mass at $T_\mathrm{release}=10$ orbits (violet curve, run $\mathcal{M}_{10}(\beta_5)$) or waiting until a deep gap has been carved at $T_\mathrm{release}=200$ orbits (green curve, run $\mathcal{M}_{200}(\beta_5)$). For $\mathcal{M}_{10}(\beta_5)$ and before the gap is carved, the migration is fast and inwards for $\simeq200$ orbits. The eccentricity increases for $\simeq40$ orbits, before being quickly damped by first-order co-orbital resonances, probably due to the high density in the co-orbital region. Once the gap is formed (after $200$ orbits), we find in both runs that there is almost no migration and the eccentricity slowly increases with a similar growth rate. We find as expected that the gap density is much lower when decreasing the magnetization (as already seen in simulations with fixed planets in \citetalias{Wafflard-Fernandez&Lesur2023}), which suggests that first-order co-orbital resonances should not be able to damp the eccentricity. Moreover, Fig.~\ref{fig:main_migration_eccentricity_b5} shows that $e_{\rm{p}}$ still increases at $\beta_0=10^5$, but with a much lower growth rate and a smaller variability compared to the case with $\beta_0=10^3$ in Fig.~\ref{fig:main_migration_eccentricity}. We suspect that accretion of matter continuously occurs in the gap with a slight gap asymmetry, enhancing first-order Lindblad resonances, and therefore the eccentricity. In consequence, a stronger wind-driven accretion seems to favor eccentricity excitation via finite amplitude instability. The influence of the disk magnetization has yet to be confirmed via dedicated simulations, but their computational cost makes the exploration of parameter space prohibitive.


\section{Summary and discussion}
\label{sec:conclusion}

In this numerical study, we performed with the \labelization{IDEFIX} code three 3D non-ideal MHD simulations of a migrating Jovian planet in a wind-launching magnetized disk, focusing in particular on the impact of the planet gap shape. When $t<T_\mathrm{release}$, and for an initial plasma parameter $\beta_0=10^3$ and a constant aspect ratio $h_0=0.05$, we retrieve the main results of \citetalias{Wafflard-Fernandez&Lesur2023} regarding a planet with $q_{\rm{p}}=10^{-3}$ on a fixed circular orbit in magnetized disks. In particular, the Jovian planet carves a radially asymmetric gap, which coexists with self-organized structures. Gas material crosses the gap via sporadic episodes of accretion and poloidal magnetic field is efficiently accumulated in this low density region, increasing locally the magnetic wind torque. Such enhanced magnetic wind in a planet-carved gap is compatible with the detection by \cite{Galloway-Sprietsma2023} of coherent upward flows arising from an annular gap in the AS 209 disk, near the location of a circumplanetary disk (CPD) candidate \citep{Bae2022}. Our main findings are:
\begin{enumerate}
    \item In all cases, planet migration is slow and mainly outward once a deep gap has been carved. The corresponding migration speed is a few tens of au.Myr$^{-1}$, for a planet initially in the $5-100$ au range. Even if the planet's semi-major axis increases on average, migration is still variable, with episodes of stalled migration and sometimes inward migration. Without a deep gap, the migration is initially faster and inwards until the amplitude of the corotation torque decreases abruptly, leading to a migration reversal and eventually a slow outward migration.
    \item By delimiting the region enclosing the non-steady gap and examining the $5$-orbit averaged gravitational torques, we show that the outer Lindblad torque $\langle\Gamma_{\rm{Lo}}\rangle_{\rm{5orb}}$ is negative with a decreasing amplitude with time, probably related to the typical timescale of the gap's outward erosion being shorter than the typical timescale of the planet's outward migration.
    \item The inner Linblad torque $\langle\Gamma_{\rm{Li}}\rangle_{\rm{5orb}}$ is a source of outward migration and both its global evolution and its variability have a strong influence on the total torque exerted on the Jovian planet, and therefore on its direction and amplitude of migration. The spikes of the instantaneous $\Gamma_{\rm{Li}}$ are partly due to an episodically denser inner ring leading to sporadic vortex formation through RWI at this location.
    \item The torque in the corotation region $\langle\Gamma_{\rm{C}}\rangle_{\rm{5orb}}$ is always negative due to (i) the over-density of material captured in the outer planet wake that tends to be carried inwards by the accretion flow; (ii) the negative density gradient at the planet location due to the gap's radial asymmetry. At later times , this density gradient could enhance negatively $\langle\Gamma_{\rm{C}}\rangle_{\rm{5orb}}$ and counterbalance the mean positive value of $\langle\Gamma_{\rm{Li}}\rangle_{\rm{5orb}}$, leading to episodes of stalled migration.
    \item In most cases, the planet is found to be eccentric, with two main phases: $e_{\rm{p}}$ is low ($\lesssim1.5\%$) initially, and eventually increases to moderate values ($\simeq7\%$). Such eccentricity evolution is consistent with a finite-amplitude instability: (i) first-order corotation resonances damp any wind-driven eccentricity perturbation to a circular orbit ($e_{\rm{p}}\ll1\%$), but saturate quickly at slightly higher eccentricities; (ii) first-order coorbital Lindblad resonances take over and stabilize small eccentricities ($e_{\rm{p}}\lesssim1.5\%$), possibly due to the increase in the gap's surface density with eccentricity; (iii) first-order external Lindblad resonances dominate and quickly excite moderate eccentricities ($e_{\rm{p}}\gtrsim2\%$), but are limited by the increasingly supersonic planet's epicyclic motion when $e_{\rm{p}}\simeq6-8\%$.
    \item The accretion rate in the gap is modulated by the planet eccentricity, with alternating phases of low and higher accretion, as well as sonic waves propagating from the planet towards the outer disk. The wind mass loss rate at high altitudes also appears to be modulated by the planet eccentricity, and could originate from the upper accretion layer crossing the disk's inner region, just inside the gap's inner edge.
\end{enumerate}

\noindent These results depart from the classical turbulent disk models. It is therefore necessary to build a new planet migration theory dedicated to the wind-driven accretion paradigm, as the notions developed in the classical viscous migration theory do not apply straightforwardly. An important next step for this 3D non-ideal MHD study would be to develop a simple framework for wind-driven accretion in 2D planet-disk simulations. In such a model, both the wind torque and mass loss rate would depend on the magnetic field strength and, consequently, on the distribution of magnetic flux, as proposed in \cite{Lesur2021}. However, gaps -- whether formed by planets or self-organization -- efficiently accumulate magnetic flux. Therefore, a predictive theory must account for magnetic flux transport and its accumulation in gaps. Currently, no such theory exists, leaving 2D simulations without a proper "effective" model.

\subsection{Vortex-driven migration with laminar accretion flows}
\label{sec:vortex_driven}

\cite{Lega2021,Lega2022} obtained a two-phased migration of Jovian planets in 3D low-viscosity disks with prescribed wind-driven angular momentum removal. In the first phase, a vortex develops at the gap's outer edge, driving an inward migration, and eventually disappears. In the second phase, inward migration accelerates as accretion is increasingly blocked by the planet, with the fastest migration occurring when gas material piles up at the gap’s outer edge, and slower migration when accretion is unaffected by the planet gap. Our results are similar regarding the underlying mechanisms at play, with both an effect of a vortex and the accretion behavior, but reversed. Wind-driven accretion is actually not blocked by the gap, and is on average slightly decreased (increased) at the inner (outer) edge of the gap, leading to a gas pile-up (depletion) at that location. Such gas depletion impedes any inward migration, whereas the inner gas pile-up drives an outward migration. As the inner ring is enhanced by sporadic episodes of accretion, it eventually becomes unstable to the RWI, episodically forming a decaying vortex for a few tens of orbits.

\subsection{Long-term evolution}
\label{sec:long}

We have seen through the paper that the radial asymmetry of the planet gap at high magnetization has a strong influence on the planet's orbital properties. However, the longest migration phase presented here lasted $300$ orbits, which corresponds to $T_\mathrm{migration}\simeq10$ kyr for a planet initially at $10$ au. Such value is a few hundred times smaller than the typical lifetime of a circumstellar disk, and raises the question of the long-term evolution of planet-disk-wind interactions. In particular, the radial drift of the outer gap edge has a typical velocity of $\simeq160$ au.Myr$^{-1}$ for $\mathcal{M}_{300}$ and $\simeq260$ au.Myr$^{-1}$ for $\mathcal{M}_{200}$, which is typically $\simeq2.5$ times faster than the typical migration rate of the planet in both cases (see Section~\ref{sec:planet_migration_eccentricity}). Depending on observational constraints and to limit this efficient expansion effect, a lower magnetization could decrease the typical speed of the gap drift. For $\beta_0=10^3$ and if the gap expansion remains the same after a few million years, we could end up with a single planet that has created a wide cavity of its own. Results from viscous simulations of disk-planet interactions in large inner cavities show that the eccentricity of massive planets is strongly increased ($\simeq0.4$) by first-order external Lindblad resonances \citep[see, e.g.,][and Section~\ref{sec:main_eccentricity}]{Papaloizou2001,Debras2021}. We might therefore expect a similar eccentricity amplification for long-time planet-disk-wind simulations, but with an eccentric outward migration rather than an inward migration, and with observational signatures similar to those studied in \cite{Baruteau2021}.

\subsection{Eccentric planets and structured winds}
\label{sec:structured_wind}

\cite{deValon2022} observed discrete structures (arches, cusps, fingers) in the rotating conical CO outflow of the DG Tau B system. They argue that the morphology and kinematics of this structured outflow is consistent with a steady MHD disk wind perturbed by quasi periodic brightness enhancements. They also show that the typical timescale between these enhancements is $T_\chi\in[190-490]$ years \citep[see the right panel of Fig.~14 in][]{deValon2020}. Assuming that this typical period is linked to the epicyclic frequency $\kappa_{\rm{p}} = 2\pi/T_\chi$ of an eccentric giant planet modulating in a periodic way the ejection rate, we can simply use Kepler's third law to retrieve the putative location $a_\chi$ of the eccentric perturber in the disk,
which gives $a_\chi\in[33-62]$ au for DG Tau B. We note that \cite{deValon2020} detected a dark ring in the radio emission of DG Tau B at $r_{\rm{dip}}\simeq53\pm5$ au, which falls inside our numerical estimate. Note that in order to establish a direct connection between the eccentricity-driven modulation of the mass loss rate and the wind structures in DG Tau B, we need to perform more detailed and precise computations, in particular via radiative transfer calculations that are beyond the scope of this paper. Nonetheless, an extension to this work would be to evaluate the visibility and detectability of the ejecta we observe in our simulations, but also to analyze how they propagate on vertical scales much larger than those considered here.\\

\begin{acknowledgements}
The authors acknowledge support from the European Research Council (ERC) under the European Union Horizon 2020 research and innovation program (Grant agreement No. 815559 (MHDiscs)). This work was granted access to the HPC resources of IDRIS under the allocations A0140402231 and A0160402231 made by GENCI. The 3D simulations were performed with the versions \href{https://github.com/idefix-code/idefix/releases/tag/v1.1}{v1.1} and \href{https://github.com/idefix-code/idefix/releases/tag/v2.0.04}{2.0.04} of \labelization{IDEFIX}. Some of the computations presented in this paper were performed using the \href{https://gricad.univ-grenoble-alpes.fr}{GRICAD} infrastructure, which is supported by Grenoble research communities. The data presented in this work were processed and plotted with Python via various libraries, in particular \href{https://github.com/numpy/numpy}{\labelization{numpy}}, \href{https://github.com/matplotlib/matplotlib}{\labelization{matplotlib}}, \href{https://github.com/scipy/scipy}{\labelization{scipy}} and \href{https://github.com/1313e/CMasher}{\labelization{cmasher}}, but also the following personal projects under development: \href{https://github.com/volodia99/nonos}{\labelization{nonos}} to analyze results from \labelization{IDEFIX} simulations, \href{https://github.com/volodia99/cblind}{\labelization{cblind}} for most of the colormaps and \href{https://github.com/volodia99/lick}{\labelization{lick}} for the visualizations in line integral convolution. IA tools have occasionally been used for debugging and reformulation. G.W.F. wishes to thank Clément Baruteau for fruitful scientific discussions, and Clément Robert for useful insights on code maintenance and development.
\end{acknowledgements}

\bibliographystyle{aa}
\bibliography{Wafflard-Fernandez_Lesur2025}

\onecolumn 
\begin{appendix}

\section{High-frequency variability of the inner Linblad torque}
\label{sec:appendix_fft}

\begin{figure}[H]
    \centering
    \includegraphics[width=0.99\hsize]{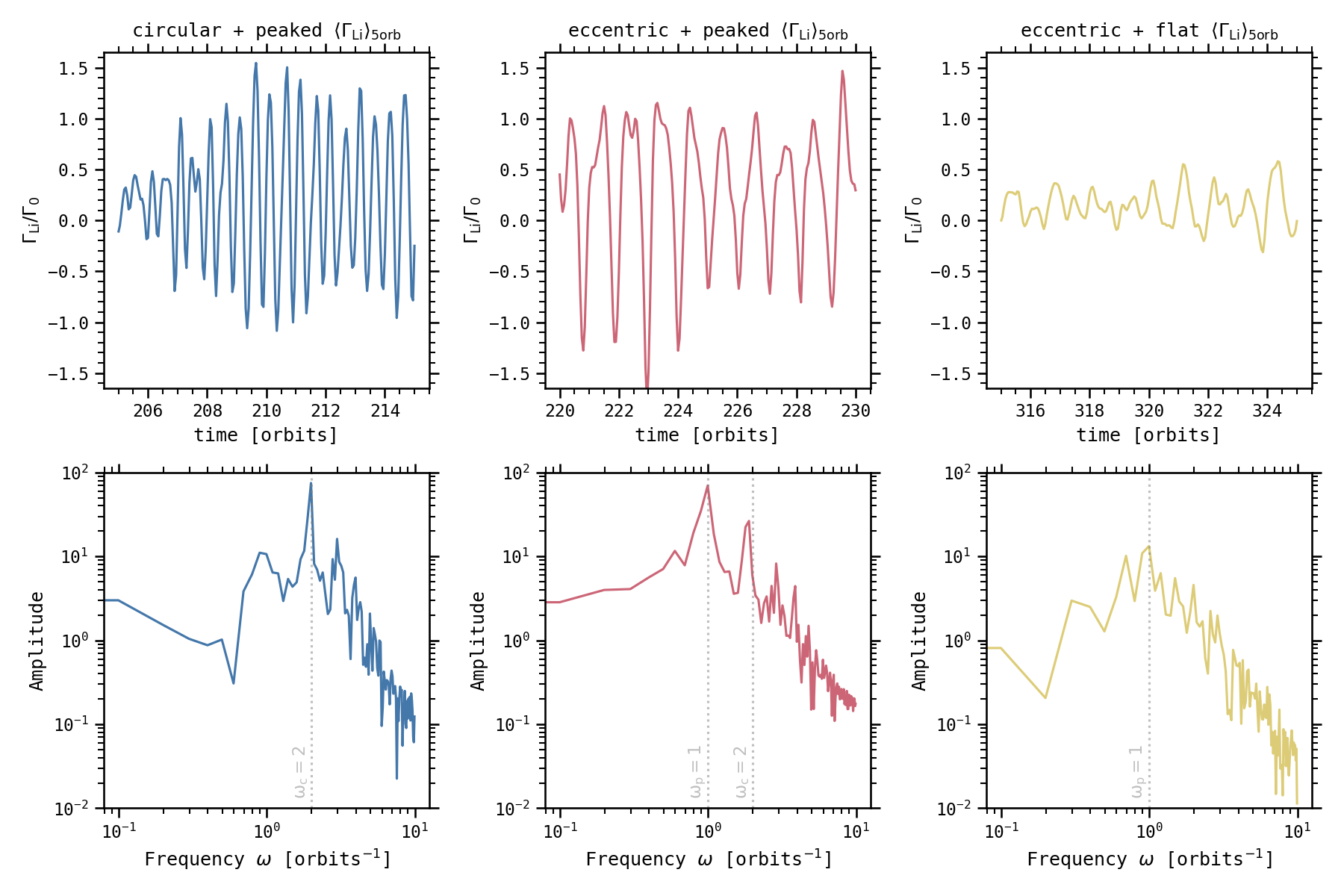}\
    \caption{$10$-orbit temporal evolution (top panels) and power-spectrum (bottom panels) of the instantaneous inner Lindblad torque $\Gamma_{\rm{Li}}$, for three epochs in the run $\mathcal{M}_{200}$. Left panels: the planet is in a quasi-circular orbit and the $5$-orbit averaged torque $\langle\Gamma_{\rm{Li}}\rangle_{\rm{5orb}}$ is peaking in Fig.~\ref{fig:main_torque}. Middle panels: the planet orbit becomes eccentric and $\langle\Gamma_{\rm{Li}}\rangle_{\rm{5orb}}$ is peaking. Right panels: the planet orbit is eccentric and $\langle\Gamma_{\rm{Li}}\rangle_{\rm{5orb}}$ does not vary. Depending on the epoch, we retrieve two typical frequencies $\rm{\omega_p}\simeq1$ and $\rm{\omega_c}\simeq2$ (vertical dotted lines).}
    \label{fig:appendix_fft}
\end{figure}

In the top panels of Fig.~\ref{fig:appendix_fft}, we present the temporal evolution of the instantaneous inner Lindblad torque $\Gamma_{\rm{Li}}$ normalized by $\Gamma_0$, for three epochs in the run $\mathcal{M}_{200}$. Using a Fourier
analysis, we obtain in the bottom panels of Fig.~\ref{fig:appendix_fft} the power spectra of $\Gamma_{\rm{Li}}$ for the three epochs of interest. Depending on the epoch, we show that the two typical frequencies $\rm{\omega_p}\simeq1$ and $\rm{\omega_c}\simeq2$ that appear in the power spectra are consistent with an eccentric forcing of the planet and a strong $m=1$ vortex sporadically developing at the location of the inner density ring :

\begin{itemize}
    \item For $t\in[200-210]$ orbits (left panels of Fig.~\ref{fig:appendix_fft}), the planet is in a quasi-circular orbit and $\rm{\omega_p}$ is not clearly visible. On the contrary, $\rm{\omega_c}$ is dominant while the $5$-orbit averaged $\langle\Gamma_{\rm{Li}}\rangle_{\rm{5orb}}$ is peaking in the middle panel of Fig.~\ref{fig:main_torque}.
    \item For $t\in[220-240]$ orbits (middle panels of Fig.~\ref{fig:appendix_fft}), $\rm{\omega_p}$ and $\rm{\omega_c}$ are both present in the power spectrum, while both the planet eccentricity and $\langle\Gamma_{\rm{Li}}\rangle_{\rm{5orb}}$ strongly increase.
    \item For $t\in[300-325]$ orbits (right panels of Fig.~\ref{fig:appendix_fft}), the planet is eccentric while $\langle\Gamma_{\rm{Li}}\rangle_{\rm{5orb}}$ is small and constant. During this epoch, the power spectrum of $\Gamma_{\rm{Li}}$ indicates indeed that $\rm{\omega_p}$ is the dominant source of high-frequency variability.
\end{itemize}

\newpage

\section{Flow in the horseshoe region}
\label{sec:appendix_horseshoe}

\begin{figure}[H]
    \centering
    \includegraphics[width=0.99\hsize]{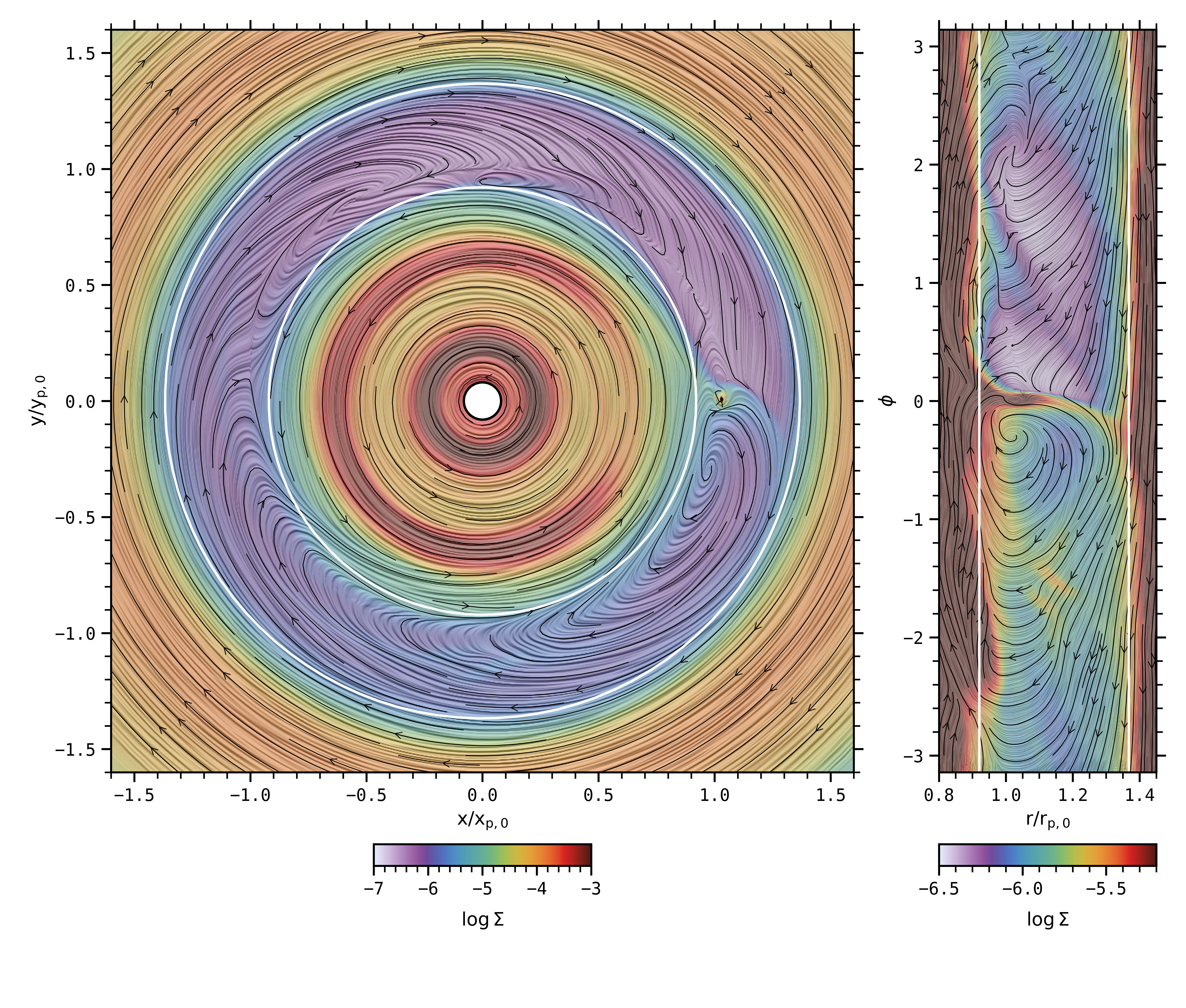}\
    \vspace{-1.2cm}
    \caption{Gas flows around the planet for the run $\mathcal{M}_{300}$ at $t=490.75$ orbits. The background color represents the logarithm of the gas surface density. The black streamlines and the LIC (Line Integral Convolution) correspond to the radial and azimuthal components of the horizontal velocity, in the midplane and in the planet’s reference frame. The white lines indicate the estimated inner and outer gap edge. The right panel focuses on the horseshoe region.}
    \label{fig:appendix_horseshoe}
\end{figure}

We plot in the left panel Fig.~\ref{fig:appendix_horseshoe} the structure of the flow around the planet for the run $\mathcal{M}_{300}$ at $t=490.75$ orbits, in a cartesian plane ($x/x_{\rm{p,0}}$,$y/y_{\rm{p,0}}$) with $x_{\rm{p,0}}^2+y_{\rm{p,0}}^2=r_{\rm{p,0}}^2$. The planet has carved a deep annular gap, and we focus on this region in the right panel of Fig.~\ref{fig:appendix_horseshoe}. We note the presence of the strong elongated vortex near the gap's inner edge, which is a source of variability of $\langle\Gamma_{\rm{Li}}\rangle_{\rm{5orb}}$. Not only is the gap radially asymmetric, it is also azimuthally asymmetric. Let us define the azimuth of the planet $\phi_{\rm{p}}$. The horseshoe region on the trailing side ($\phi<\phi_{\rm{p}}$) of the planet is reduced and is on average subject to a mass excess, unlike the region on the leading side ($\phi>\phi_{\rm{p}}$) of the planet which is subject on average to a mass deficit. This imbalance may be responsible for a fraction of the negative corotation torque. This gas dynamics we observe in the gap region is in fact the opposite of the prediction by \cite{Kimmig2020}, and can be interpreted the following way. Because the horseshoe orbits are disrupted by the accretion flow, some material is able to perform horseshoe U-turns on the trailing side close to the planet, but along the U-turn the material is progressively pushed inwards by the accretion flow until it reaches the inner gap edge, no further than an azimuth $\phi\simeq\phi_{\rm{p}}-\pi/2$ on average, and therefore well before reaching the leading side close to the planet at $\phi_{\rm{p}}-2\pi$. This strong azimuthal asymmetry of the horseshoe region makes the corotation torque unsaturated 
and negative due to the accumulation of material close to the trailing side of the planet, always replenished by the radial accretion flow 
. In particular, the over-density of material captured in the outer planet wake tends to be carried away by the accretion flow (see over-dense material at $(r,\phi)\simeq (1.15r_{\rm{p,0}},\phi_{\rm{p}}-\pi/2)$), crossing the gap beneath the planet in azimuth and reaching the inner gap edge afterwards.

\newpage

\section{Eccentricity-driven ejection modulation at two opposite surfaces}
\label{sec:appendix_modulation}

\begin{figure}[H]
    \centering
    \includegraphics[width=0.99\hsize]{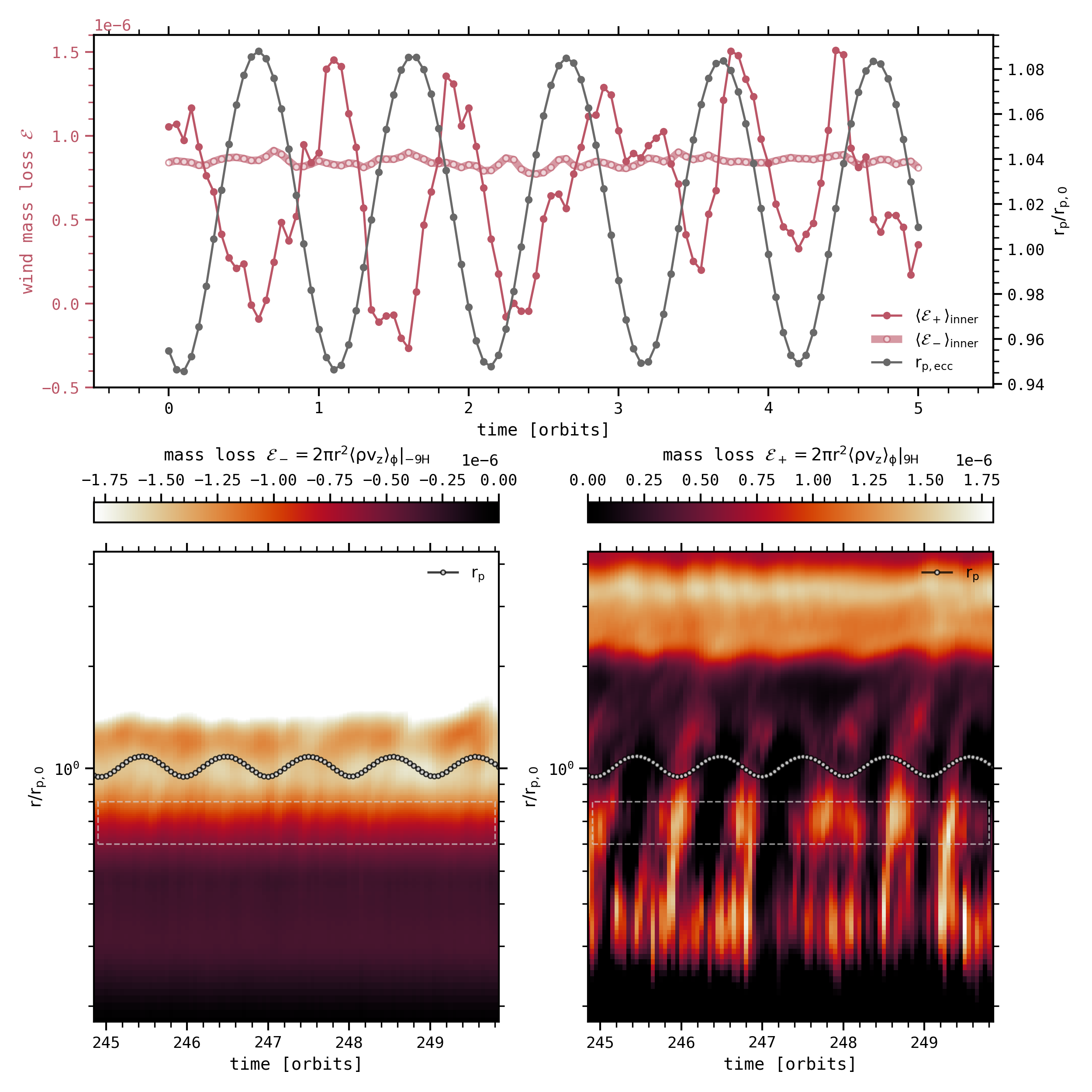}\
    \caption{Eccentricity-driven modulation of the ejection processes at two surfaces $\theta=\Theta_+=\arctan(9h)$ and $\theta=\Theta_-=-\arctan(9h)$ for the run $\mathcal{M}_{200}$. Top panel: $5$-orbit evolution of the planet radius $r_{\rm{p}}$ (gray line) and of the wind mass loss rate, averaged over the radial interval $[0.6r_{\rm{p,0}}-0.8r_{\rm{p,0}}]$ and estimated at $\Theta_+$ (thin red line) and $\Theta_-$ (thick red line). Bottom panels: $5$-orbit space-time diagrams of $\mathcal{E}_-$ (left panel) and $\mathcal{E}_+$ (right panel). The instantaneous radial position of the planet is indicated with white circles, and the radial interval used to average $\mathcal{E}$ in the top panel is indicated by white dashed rectangles.}
    \label{fig:appendix_modulation}
\end{figure}

In Fig.~\ref{fig:appendix_modulation}, we plot the temporal evolution over $5$ orbits of the wind mass loss rate $\mathcal{E}_+$ and $\mathcal{E}_-$ at two opposite latitudes far from the disk surfaces, respectively at $\Theta_+=\arctan(9h)\gg\theta_+=\arctan(3h)$ and $\Theta_-=-\Theta_+$. The top panel of Fig.~\ref{fig:appendix_modulation} shows $\mathcal{E}_+$ (thin red line) and $\mathcal{E}_-$ (thick red line) integrated on the radial interval $[0.6r_{\rm{p,0}}-0.8r_{\rm{p,0}}]$, thus coming from the inner disk region. The bottom panels of Fig.~\ref{fig:appendix_modulation} show the spacetime diagrams for $\mathcal{E}_+$ (left) and $\mathcal{E}_-$ (right), obtained by performing a radial average over a moving radial band with a width of $0.2r_{\rm{p,0}}$. We retrieve an eccentricity-driven modulation of $\mathcal{E}_+$ which seems to originate from an upper accretion layer also modulated by the planet eccentricity. In contrast, the profile of $\mathcal{E}_-$ remains relatively flat over time.

\end{appendix}

\end{document}